\documentclass[preprintnumbers]{revtex4}
\usepackage{amsfonts}
\usepackage{amssymb}
\usepackage{amsmath}
\usepackage{epsfig}
\usepackage{tabularx}
\usepackage{array}
\allowdisplaybreaks
\begin{document}
\title{Collapsing dynamics of relativistic fluid in modified gravity admitting a conformal Killing vector}





\title{\textbf{Collapsing dynamics of relativistic fluid in modified gravity admitting a conformal Killing vector}}
\author{Kazuharu Bamba$^1$
\thanks{bamba@sss.fukushima-u.ac.jp}, Z. Yousaf$^2$ \thanks{zeeshan.math@pu.edu.pk}, M. Z. Bhatti$^2$
\thanks{mzaeem.math@pu.edu.pk} and R.
Nazer$^2$ \thanks{rukhsananazer8@gmail.com}, and Yuki Hashimoto$^1$ \thanks{s2270040@ipc.fukushima-u.ac.jp}\\
$^1$ Faculty of Symbiotic Systems Science, \\ 
Fukushima University, Fukushima 960-1296, Japan\\
$^2$ Department of Mathematics, University of the Punjab, \\
Quaid-i-Azam Campus, Lahore-54590, Pakistan}


\begin{abstract}

The collapsing dynamics of relativistic fluid are explored in $f(R)$ gravity in a detailed systematic manner for the non-static spherically symmetric spacetime satisfying the equation of the conformal Killing vector. With quasi-homologous condition and diminishing complexity factor condition, exact solutions for dissipative as well as for non-dissipative systems are found and the astrophysical applications of these exact solutions are discussed. Furthermore, it is demonstrated that $f(R)=R$, which is the extensive restriction of $f(R)$ gravity, prior solutions of the collapsing fluid in general relativity, can be retrieved.

\end{abstract}
\maketitle

\section{introduction}

The evolution of the universe has always been an interesting topic for scientists. Different models were proposed by different scientists to explain the initial state, evolution, and final fate of our universe. After a tremendous explosion referred to as the ``Big Bang" \cite{lemaitre2013republication}, our universe began as an incredibly small dot that quickly began to expand due to its extreme density and heat. Many physicists accepted that the Big Bang theory, which describes how the universe grew from a denser, hotter childhood, accurately captured the nature of the universe \cite{hubble1929relation, penzias1965measurement, peebles1991case}. The celestial objects, including planets, all shining matter, and stars, are made of matter (referred to as baryonic matter), which constitutes a small part of this dark cosmos, and another minor contributor is electromagnetic radiation. Dark energy, which has controlled the current energy balance, is more frequent than dark matter, which has not yet been found in a laboratory.

In the same way, celestial objects (stars, planets, and galaxies) also undergo changes in their life cycles. The structure and formation of celestial objects in the universe are a result of gravitational collapse \cite{penrose2002golden}, which is a marvelous issue in the theory of gravitation and relativistic astrophysics and attracted the attention of researchers. The factor that balances the star is the equilibrium between the outward-directed pressure and the gravitational pull that directs it inward. In gravitational collapse, the celestial object contracts towards its center under the dominant action of its own gravity. The result of this collapse is the emergence of compact (dense) objects that includes black holes, white dwarfs, and neutron stars. If the star has a mass that is just a few times that of the sun, it consumes its nuclear fuel and collapses. According to the general relativity theory (GR), the result of this collapse is either a black hole or a naked singularity. Spherically symmetric spacetime has been taken into consideration by many researchers when discussing the majority of gravitational collapse-related issues.

Oppenheimer and Snyder \cite{oppenheimer1939continued} studied the collapse of dust. They found the solutions of Einstein's field equations (EFEs) describing this process. Friedman and Schutz \cite{friedman1978secular} looked into how gravitational radiation and viscosity affected the evolution of stars. In some cases, viscosity keeps rotating stars stable, whereas gravitational radiation makes them unstable. Lake and Hellaby \cite{lake1981collapse} confirmed the presence of the naked singularity after examining the radiating spherical collapse. Joshi and Singh \cite{joshi1995role} studied the spherically symmetric dust collapse encapsulating irregular distribution of energy density. They came to the conclusion that the  energy density of a star and its radius has a significant impact on the outcome. The inhomogeneity factor that may result in naked singularities was identified by Herrera et al. \cite{herrera1998role}. Chan \cite{chan2000radiating} analyzed the collapse of the relativistic star model and figured out that the shear viscosity of the fluid contributes to the  increase in anisotropic pressure. Wang \cite{wang2003critical} in his investigation of the cylindrical collapse,  discovered a limitation that may cause a collapse to produce black holes. Herrera et al. \cite{herrera2004spherically} discovered a link between anisotropic pressure, shear, and Weyl scalars and found a constraint on the density irregularity of a diffusing star. Sharif and Yousaf \cite{sharif2015radiating} investigated the collapse of a perfect fluid. They  investigated that the involvement of the constant curvature terms slows down the rate of collapse. Cipolletta and Giamb\'{o} \cite{cipolletta2012collapse} studied the effects of electromagnetic field on the gravitational collapse of the spherical relativistic geometries. It turns out that in this anisotropic scenario, electromagnetic charge entirely altered the final state of the imploding star.

The GR is a successful theory of cosmology that agrees with many solar system tests in the presence of a weak gravitational field. At large scales or in strong gravitational fields, this theory may need some modifications to unveil various hidden cosmological aspects, like the dark energy problem. The modifications in GR can help to explain the cosmic accelerating expansion. These modifications are referred to as modified gravity theories (MGTs) \cite{padmanabhan2008dark,lee2019motion,clifton2012modified,bamba2012cosmic,capozziello2011extended,nojiri2011unified,sotiriou2010f,de2010f,joyce2015beyond,bamba2015inflationary,nojiri2017modified}. The MGTs include $f(R)$ \cite{nojiri2007unifying,olmo2011palatini,capozziello2002curvature}, $f(G)$ \cite{nojiri2005modified, bamba2017energy}, $f(R, T)$ \cite{harko2011f,yousaf2016causes} (where $G$ represents Gauss Bonnet-invariant, the symbol $R$ stands for Ricci-scalar and the quantity $T$ symbolizes trace of stress-energy tensor), etc. Out of these MGTs, the simplest one is the $f(R)$ theory. The idea of $f(R)$ theory has fascinated great attention because it may provide the simplest description of DE. This theory is obtained when we have replaced the Ricci scalar $R$ by a generic function of $R$ i.e., $f(R)$, in the Einstein-Hilbert action. The $f(R)$ gravity is presented by Nojiri and Odintsov \cite{nojiri2003modified} and they analyzed that this theory can explain the accelerating expansion of the universe in an effective way.

Many researchers investigated the spherical star collapsing phenomenon using various fluid configurations in $f(R)$ gravity. Sharif and Kausar \cite{sharif2011gravitational} analyzed the collapse of the perfect fluid. They deduced that current curvature corrections produce repulsion effects to reduce the collapse rate. Sebastiani et al. \cite{sebastiani2013instabilities} used a special $f(R)$ formalism to study the formulation of Nariai black holes and figured out that the choice of $f(R)$ model affects the stages at which a collapsing system occurs. Astashenok et al. \cite{astashenok2013further,astashenok2015extreme} discussed a few features of stellar systems in various modified gravity models. They described some eras of model parameters that support the formation of massive stellar structures in a  better way. A set of structure scalars can be used explicitly to write all static anisotropic cylindrical solutions \cite{herrera2012cylindrically}. Bhatti et al. \cite{bhatti2017gravitational} studied the collapse of electrically charged Lema\^{i}tre-Tolman-Bondi spacetime and evaluated the effect of $f(R)$ theory on this geometry. They investigated a group of solutions to the $f(R)$ field equations in the presence of an electromagnetic field and with a constant curvature scalar. Malik et al. \cite{malik2023investigation} studied the static non-rotating stellar models in $f(R)$ gravity. They deduced the corresponding field equations for some specific star models and discovered the connection between the mass variables and radius. Mustafa et al. \cite{mustafa2021wormhole} derived wormhole solutions by utilizing the Karmarkar condition \cite{sufyananisotropic} and described the possibility for the formation of traversable wormholes meeting the energy requirements. Oikonomou \cite{oikonomou2022f} described a to understand gravitational waves phenomenon in $f(R)$ gravity and studied the impact of the post-inflationary era. In $f(R)$ gravity, spherically symmetric static solutions associated with electromagnetic fields were generated by Nashed and Nojiri \cite{nashed2023black}. They demonstrated that the curvature singularity in GR is substantially softer due to the higher-derivative terms than it is in charged black holes. Recently, Oikonomou \cite{oikonomou2023rp,oikonomou2023static} performed powerful numerical simulations to analyze the existence of neutron stars corresponding to particular equations of state and attractors.

We require the analytical solutions of the nonlinear field equations to investigate the internal geometry of celestial objects, represented by the equation of state (EoS). The EoS gives the relationship between the pressure and the matter density. The degree of nonlinearity of EFEs is one of the major challenges in connecting the unique characteristics of GR to actual physical issues. Therefore, it becomes very challenging to solve these Eq.s without imposing specific symmetry constraints on a space-time metric. These isometries are called killing vectors (KVs) and give rise to conservational laws. In spite of the nonlinearity of these partial differential equations, numerous researchers found exact, astrophysical and cosmological solutions of field equations. The field equations for an isotropic sphere in vacuum were  first solved by Schwarzschild \cite{schwarzschild1916sitzungsber}. Rahaman et al. \cite{rahaman2014exact} studies solutions of compact stars by considering the barotropic equation of state. Malaver \cite{malaver2015charged} considered the equation of state to determine the distinct forms for the gravitational potential. Two solutions of field equations for compact objects like neutron or quark stars were presented by Zubairi et al. \cite{zubairi2015static}. In order to solve field equations for spherically symmetric mass configurations (for a limited value of the cosmological constant), they investigated the composition of distorted (non-spherical) dense objects. For various configurations of conformal Killing vectors (CKV), Herrera et al. \cite{herrera2022non} discovered a variety of exact analytical solutions, in dissipative as well as in adiabatic regimes. In order to find particular solutions, they imposed some restrictions on the system such as diminishing complexity factor \cite{herrera2019complexitya,herrera2020stability,herrera2019complexity} and quasi-homologous evolution \cite{herrera2020quasi}.

In this paper, we study spherically symmetric fluid configurations that are enclosed in a surface $(\Sigma)$. The source is an anisotropic fluid with dissipation freedom. We find a few exact, non-static solutions satisfying a one-parameter group of conformal motions. We take the two cases: either four-velocity is parallel to the killing vector field $(\chi^{\nu}\|V^{\nu})$ or four-velocity is perpendicular to the killing vector field $(\chi^{\nu}\bot V^{\nu})$. Each case is further discussed in both dissipative $(q\neq0)$ and non-dissipative $(q=0)$ regimes. To find particular solutions, we impose some restrictions, like vanishing complexity factor conditions, quasi-homologous conditions, etc. In Sec. \textbf{II}, the metric, source, and kinematical variables are defined and relevant equations are calculated. The expression of the complexity factor in the form of metric coefficients is given. In Sec. \textbf{III}, the expression of the transport equation is given, which is useful to analyze the distribution and evolution of temperature. In Sec. \textbf{IV}, the homologous and quasi-homologous conditions are defined. In Sec. \textbf{V}, we start  by assuming the line element of interior spacetime satisfying the CKV equation and we found a number of solutions of the given system. Sec. \textbf{VI} includes findings and an analysis of the physical applicability of obtained solutions. Finally, Appendices are given that include helpful formulas.

\section{The Metric, energy-momentum tensor, physical variables, and related equations}

The action of $f(R)$ gravity minimally coupled with matter Lagrangian density $\mathbb{L}_{m}$ is described as
\begin{align}\label{1}
\mathbb{A}=\int\sqrt{-g}d^{4}x\bigg[\frac{f(R)}{2\kappa}+\mathbb{L}_{m}(\varphi_{m},g_{\rho\nu})\bigg],
\end{align}
where $\kappa$ is a coupling constant, the symbol $\varphi_{m}$ specifies the matter field and the scalar $g$ is the determinant of the metric tensor $g_{\rho\nu}$.
Upon varying Eq.\eqref{1} in regard to $g_{\rho\nu}$, fourth-order partial differential equations are obtained as under
\begin{align}\label{2}
f_{R} R_{\rho\nu}-\frac{1}{2}f(R)g_{\rho\nu}-\nabla_{\rho}\nabla_{\nu}f_{R}+
g_{\rho\nu}\nabla^{a}\nabla_{a}f_{R}=8\pi G\mathbb{T^{(\mathfrak{M})}_{\rho\nu}},
\end{align}
where $f_{R}\equiv\frac{df}{dR}$, $R_{\rho\nu}$ is the Ricci tensor, the notation of covariant derivative is $\nabla_{\rho}$ and $\mathbb{T^{(\mathfrak{M})}_{\rho\nu}}$ is the energy-momentum tensor and is given by
\begin{align}\label{3}
\mathbb{T^{(\mathfrak{M})}_{\rho\nu}}=\rho V_{\rho}V_{\nu}+Ph_{\rho\nu}+\Pi_{\rho\nu}+q(V_{\rho}K_{\nu}
+K_{\rho}V_{\nu}),
\end{align}
with
\begin{align*}
P=\frac{P_{r}+2P_{\bot}}{3},\quad\quad h_{\rho\nu}=g_{\rho\nu}+V_{\rho}V_{\nu},\quad\quad \Pi_{\rho\nu}=\Pi\bigg(K_{\rho}K_{\nu}-\frac{1}{3}h_{\rho\nu}\bigg), \quad\quad \Pi=P_{r}-P_{\bot},
\end{align*}
where $\mu$ is energy density, $P_{r}$ and $P_{\bot}$ are stress components along radial and tangential directions, respectively.  $V^{\rho}$, $q^{\rho}$, and $K^{\rho}$ are four velocities, the heat flux, and a unit four vector along the radial direction, respectively. Equation \eqref{2} can be rearranged as
\begin{align}\label{4}
G_{\rho\nu}=\frac{8\pi}{f_{R}}\bigg(\mathbb{T^{(\mathfrak{M})}_{\rho\nu}}+
\mathbb{T^{(\mathfrak{D})}_{\rho\nu}}\bigg),
\end{align}
where
\begin{align}\label{5}
\mathbb{T^{(\mathfrak{D})}_{\rho\nu}}=\frac{1}{8\pi}\bigg[\frac{f-Rf_{R}}{2}g_{\rho\nu}+\nabla_{\rho}
\nabla_{\nu}f_{R}-g_{\rho\nu}\Box f_{R}\bigg],
\end{align}
where $\Box$ is the d'Alembert operator. We model our system in such a way that the geometry of the interior region is described through spherically symmetric line element as
\begin{align}\label{6}
ds^2=-A^{2}dt^{2}+B^{2}dr^{2}+C^{2}(d\theta^{2}+\sin^{2}\theta d\phi^{2}),
\end{align}
where $A$, $B$ and $C$ depends upon the coordinates $t$ and $r$ and are positive. Here the
coordinates are represented as $\mathbf{x}^{0}=t$, $\mathbf{x}^{1}=r$, $\mathbf{x}^{2}=\theta$ and $\mathbf{x}^{4}=\phi$. Also, $A$ and $B$ have no dimension, while the dimension of $R$ is $length$. Since we have assumed that the observers are co-moving, we have
\begin{align}\label{7}
V^{\rho}=A^{-1}\delta^{\rho}_{0},\quad\quad q^{\rho}=qK^{\rho}, \quad\quad K^{\rho}=B^{-1}\delta^{\rho}_{1}.
\end{align}
These terms fulfill the identities
\begin{align}\label{8}
V^{\rho}V_{\rho}=-1, \quad\quad V^{\rho}q_{\rho}=0,\quad\quad K^{\rho}K_{\rho}=1,\quad\quad K^{\rho}V_{\rho}=0.
\end{align}
The acceleration, expansion scalar, and shear of the fluid are given, respectively, as
\begin{align}\label{9}
a_{\rho}=V_{\rho;\nu}V^{\nu},\quad\quad \Theta=V^{\rho}_{;\rho},\quad\quad
\sigma_{\rho\nu}=V_{(\rho;\nu)}+a(\rho V_{\nu})-\frac{1}{3}\Theta h_{\rho\nu}.
\end{align}
From the above expressions, we can write
\begin{align}\label{10}
a_{\rho}=aK_{\rho},\quad\quad a=\frac{A'}{AB},\quad\quad
\Theta=\frac{1}{A}\bigg(\frac{\dot{B}}{B}+2\frac{\dot{C}}{C}\bigg),
\end{align}
here the notations dot and prime show that the derivatives are evaluated with regard to $t$ and $r$, respectively. The nonvanishing components of shear are
\begin{align}\label{11}
\sigma_{11}=\frac{2}{3}B^{2}\sigma, \quad\quad \sigma_{22}=\frac{\sigma_{33}}{\sin^{2}\theta}=-\frac{1}{3}C^{2}\sigma,
\end{align}
where $\sigma^{\rho\nu}\sigma_{\rho\nu}=\frac{2}{3}\sigma^{2}$, with
\begin{align}\label{12}
\sigma=\frac{1}{A}\bigg(\frac{\dot{B}}{B}-\frac{\dot{C}}{C}\bigg),
\end{align}
is the shear scalar. The expression for Misner-Sharp mass \cite{misner1964relativistic} is
\begin{align}\label{13}
m=\frac{C^{3}}{2}{R_{23}}^{23}=\frac{C}{2}\bigg[\bigg(\frac{\dot{C}}{A}\bigg)^{2}-
\bigg(\frac{C'}{B}\bigg)^{2}+1\bigg].
\end{align}
The proper time derivative $D_{T}$ is defined as $D_{T}=\frac{1}{A}\frac{\partial}{\partial t}$
Through this operator, we define the velocity $\mathbb{V}$ of the fluid, interpreted as
$\mathbb{V}=D_{T}C$, which is assumed to be negative here. Thus, the Eq. \eqref{13} takes the form
\begin{align}\label{16}
E\equiv\frac{C'}{B}=\bigg(1+\mathbb{V}^{2}-\frac{2m}{C}\bigg)^\frac{1}{2}.
\end{align}
Utilizing Eq. \eqref{16}, \eqref{eq:A6} can be written as
\begin{align}\label{17}
\frac{4\pi}{f_{R}}\bigg[qAB-\frac{1}{8\pi}\bigg(\dot{f'_{R}}-\frac {A'}{A}\dot{f_{R}}-\frac{\dot{B}}{B}f'_{R}
\bigg)\bigg]=AC'\bigg[\frac{1}{3}D_{R}(\Theta-\sigma)-\frac{\sigma}{C}\bigg],
\end{align}
with
\begin{align}\label{18}
D_R=\frac{1}{C'}\frac{\partial}{\partial r}.
\end{align}
The mass variation with the help of previously defined operators can be written as
\begin{align}\nonumber
D_{T}m=&\frac{4\pi}{f_{R}}(P_{r}\mathbb{V}+qE)C^{2}-\frac{C^{2}}{2Af_{R}}\bigg[\frac{\dot{C}}{2}(f-Rf_{R})+
\frac{\dot{C}}{A^{2}}\ddot{f_{R}}-\frac{C'}{B^{2}}\dot{f'_{R}}+\\\label{19}&\bigg\{\frac{\dot{C}}{A^{2}}\bigg(2\frac{\dot{C}}{C}
-\frac{\dot{A}}{A}\bigg)+\frac{C'}{B^{2}}\frac{A'}{A}\bigg\}\dot{f_{R}}-\bigg\{\frac{\dot{C}}{B^{2}}\bigg(
\frac{A'}{A}+2\frac{C'}{C}\bigg)-C'\frac{\dot{B}}{B^{3}}\bigg\}f'_{R}\bigg],
\end{align}
and
\begin{align}\nonumber
D_{R}m=&\frac{4\pi}{f_{R}}\bigg(\frac{\mathbb{V}}{E}q+\mu\bigg)C^{2}+\frac{C^{2}}{2f_{R}}\bigg[\frac{1}{B^{2}}f''_{R}
-\frac{\dot{C}}{C'A^{2}}\dot{f'_{R}}-\frac{1}{A^{2}}\bigg(\frac{\dot{C}}{C'}\frac{A'}{A}-2\frac{\dot{C}}{C}+
\frac{\dot{B}}{B}\bigg){\ddot{f_{R}}}+\\\label{20}&\bigg\{\frac{1}{B^{2}}\bigg(2\frac{C'}{C}-
\frac{B'}{B}\bigg)+\frac{\dot{C}}{C'A^{2}}\frac{\dot{B}}{B}\bigg\}f'_{R}-\frac{f-Rf_{R}}{2}\bigg].
\end{align}
Equation \eqref{22} gives
\begin{align}\label{21}
m=&4\pi\int\frac{1}{f_{R}}\bigg(\frac{\mathbb{V}}{E}q+\mu\bigg)C'C^{2}dr+\int\frac{C'C^{2}}{2f_{R}}\bigg[\frac{1}{B^{2}}f''_{R}
-\frac{\dot{C}}{C'A^{2}}\dot{f'_{R}}-\frac{1}{A^{2}}\bigg(\frac{\dot{C}}{C'}\frac{A'}{A}-2\frac{\dot{C}}{C}+
\frac{\dot{B}}{B}\bigg){\ddot{f_{R}}}+\\\nonumber&\bigg\{\frac{1}{B^{2}}\bigg(2\frac{C'}{C}-
\frac{B'}{B}\bigg)+\frac{\dot{C}}{C'A^{2}}\frac{\dot{B}}{B}\bigg\}f'_{R}-\frac{f-Rf_{R}}{2}\bigg]dr.
\end{align}
which satisfies the condition $m(0)=0$.

\subsection{The Weyl Tensor and the Complexity Factor}

It is notable \cite{herrera2018new} that the complexity function is a scalar entity related to the so-called structural scalars \cite{herrera2009structure} that is used to find the level of complexity of a particular fluid configuration. The condition of the diminishing complexity factor led to some of the solutions that will be shown in the next Sect. The magnetic component of the Weyl tensor $(\mathcal{C}^{\nu}_{\alpha\beta\mu})$ vanishes for spherically symmetric spacetime; as a result, it is described by its "electric" component $\mathbb{E}_{\rho\gamma}$ only, given by
\begin{align}\label{22}
\mathbb{E}_{\rho\gamma}=\mathcal{C}_{\rho\alpha\gamma\mu}V^{\alpha}V^{\mu},
\end{align}
whose non-zero components are
\begin{align}\label{23}
&\mathbb{E}_{11}=\frac{2}{3}B^{2}\mathcal{E},\quad\quad
\mathbb{E}_{22}=-\frac{1}{3}C^{2}\mathcal{E},\quad\quad
\mathbb{E}_{33}=\mathbb{E}_{22}\sin^{2}\theta,
\end{align}
where
\begin{align}\label{24}
\mathcal{E}=\frac{1}{2A^{2}}\bigg[\frac{\ddot{C}}{C}-\frac{\ddot{B}}{B}-\bigg(\frac{\dot{A}}{A}+\frac{\dot{C}}
{C}\bigg)\bigg(\frac{\dot{C}}{C}-\frac{\dot{B}}{B}\bigg)\bigg]+\frac{1}{2B^{2}}\bigg[\bigg(\frac{C'}{C}-\frac
{A'}{A}\bigg)\bigg(\frac{B'}{B}+\frac{C'}{C}\bigg)+\frac{A''}{A}-\frac{C''}{C}\bigg]-\frac{1}{2C^{2}}.
\end{align}
The Weyl tensor's electric part can also be written as
\begin{align}\label{25}
\mathbb{E}_{\rho\gamma}=\mathcal{E}\bigg(K_{\rho}K_{\gamma}-\frac{1}{3}h_{\rho\gamma}\bigg).
\end{align}
Introduce Weyl tensor $Y_{\rho\gamma}$ in a better way
\begin{align}\label{26}
Y_{\rho\gamma}=R_{\rho\alpha\gamma\mu}V^{\alpha}V^{\mu},
\end{align}
which is further written as follows
\begin{align}\label{27}
Y_{\rho\gamma}=\frac{1}{3}Y_{T} h_{\rho\gamma}+Y_{TF}\bigg(K_{\rho}K_{\gamma}-\frac{1}{3}h_{\rho\gamma}\bigg).
\end{align}
The trace and trace free parts of Eq. \eqref{27}, i.e., scalar functions $Y_{T}$ and $Y_{TF}$ can be written as
\begin{align}\label{28}
Y_{T}=4\pi(\mu+3P_{r}-2\Pi),\quad\quad Y_{TF}=\mathcal{E}-4\pi\Pi.
\end{align}
The scalar $Y_{TF}$ has been recognized as a complexity factor by many researchers \cite{yousaf2020influence,yousaf2020complexity,leon2022complexity,yousaf2020study,yousaf2020definition,
leon2023spherically,andrade2022stellar,maurya2022relativistic}. Finally, the complexity factor in the form of metric functions reads as
\begin{align}\label{29}
Y_{TF}=\frac{1}{B^{2}}\bigg[\frac{A''}{A}-\frac{A'}{A}\frac{B'}{B}-\frac{A'}{A}\frac{C'}{C}\bigg]
+\frac{1}{A^{2}}\bigg[\frac{\ddot{C}}{C}-\frac{\ddot{B}}{B}+\frac{\dot{A}}{A}\frac{\dot{B}}{B}-
\frac{\dot{A}}{A}\frac{\dot{C}}{C}\bigg].
\end{align}

\subsection{The exterior spacetime and matching conditions}

We assume that the fluid is enclosed by the surface $\Sigma$. To deal with such a scenario, we shall compute junction conditions \cite{bonnor1981junction}. We assume that the exterior metric to $\Sigma$ is
\begin{align}\label{30}
ds^{2}=-\bigg[1-\frac{2M(\nu)}{r}\bigg]d\nu^{2}-2drd\nu+r^{2}(d\theta^{2}+\sin^{2}\theta d\phi^{2}),
\end{align}
where $M(\nu)$ signifies the total mass and the symbol $\nu$ is used to represent retarded time. The continuity of the first and second fundamental forms across $\Sigma$ is necessary for the matching of the interior spacetime to the exterior metric, on the surface $r=r_{\Sigma}= constant$ \cite{chan1997collapse}, which provides us
\begin{align}\label{31}
 m(t,r)\overset{\Sigma}= M(\nu),
\end{align}
and
\begin{align}\label{32}
 q\overset{\Sigma}=P_{r}+\frac{1}{16\pi}(f-Rf_{R}),
\end{align}
where $\overset{\Sigma}=$ indicates that both sides of the equation are examined on $\Sigma$.

\section{The transport Equation}

In order to find the temperature distribution and evolution in the dissipative case, we require a transport equation. The general expression of the transport equation reads
\begin{align}\label{33}
 \tau h^{\rho\nu}V^{\alpha}q_{\nu;\alpha}+q^{\rho}=- \mathcal{K}h^{\rho\nu}(T_{\nu}+Ta_{\nu}-
 \frac{1}{2}\mathcal{K} T^{2}\bigg(\frac{\tau V^{\nu}}{\mathcal{K} T^{2}}\bigg)_{;\nu}q^{\rho},
\end{align}
where the notation $\mathcal{K}$ is for the thermal conductivity, and $T$ is for temperature and $\tau$ simply configures the relaxation time. If the spacetime is spherically symmetric, then there exists only one  independent component of the transport equation, which is obtained from Eq. \eqref{33}, when we contract Eq. \eqref{33} with the vector $K^{\alpha}$, giving us
\begin{align}\label{34}
\tau V^{\alpha}q_{\alpha}+q=-\mathcal{K}(K^{\alpha}T_{,\alpha}+Ta)-\frac{1}{2}\mathcal{K} T^{2}\bigg(
\frac{\tau V^{\alpha}}{\mathcal{K} T^{2}}\bigg)_{;\alpha}q.
\end{align}
When we neglect the last term, Eq. \eqref{34} reads as
\begin{align}\label{35}
\tau V^{\alpha}q_{\alpha}+q=-\mathcal{K}(K^{\alpha}T_{,\alpha}+Ta).
\end{align}
This equation is called the truncated transport equation (which is comparatively easy to solve).

\section{Homologous and the quasi-homologous condition}

We will impose the diminishing complexity factor constraint in order to evaluate models. It is also necessary to clarify what is the most basic technique that defines the collapse of the system. The notion of homologous evolution was presented in Ref.~\cite{herrera2018definition}. Thus, Eq. \eqref{eq:A3} takes the form
\begin{align}\label{36}
D_{C}\bigg(\frac{\mathbb{V}}{C}\bigg)=\frac{4\pi}{E}\bigg(q+\frac{1}{8\pi}\bigg(\dot{f'_{R}}-
\frac{A'}{A}\dot{f_{R}}-\frac{\dot{B}}{B}f'_{R}\bigg)\bigg]\bigg)+\frac{\sigma}{C},
\end{align}
whose integration gives
\begin{align}\label{37}
\mathbb{V}=\tilde{a}(t)C+C\int^{r}_{0}\bigg\{\frac{4\pi}{E}\bigg((q-\frac{1}{8\pi}\bigg(\dot{f'_{R}}
-\frac{A'}{A}\dot{f_{R}}-\frac{\dot{B}}{B}f'_{R}\bigg)\bigg]\bigg)+\frac{\sigma}{C}\bigg\}C'dr,
\end{align}
where $\tilde{a}(t)$ is an integration function, or
\begin{align}\label{38}
\mathbb{V}=\frac{\mathbb{V}_{\Sigma}}{C_{\Sigma}}C+C\int^{r_{\Sigma}}_{0}\bigg\{\frac{4\pi}{E}\bigg(q-
\frac{1}{8\pi}\bigg(\dot{f'_{R}}-\frac{A'}{A}\dot{f_{R}}-\frac{\dot{B}}{B}f'_{R}\bigg)\bigg]\bigg)+
\frac{\sigma}{C}\bigg\}C'dr.
\end{align}
If the integral terms of Eqs. \eqref{37} and \eqref{38} disappear, then we obtain
\begin{align}\label{39}
\mathbb{V}=\tilde{a}(t)C.
\end{align}
The homologous evolution may take place if the integral terms of the above equations cancel each other. Later, the homologous condition was relaxed resulting in what was termed quasi-homologous evolution. The condition in Eq. \eqref{39} suggests
\begin{align}\label{40}
\frac{4\pi}{C'}B\bigg(q-\frac{1}{8\pi}\bigg[\dot{f'_{R}}-\frac{A'}{A}\dot{f_{R}}-
\frac{\dot{B}}{B}f'_{R}\bigg]\bigg)+\frac{\sigma}{C}=0.
\end{align}
This equation informs us about the simplest evolution of the fluid in metric $f(R)$ gravity.

\section{Exact solutions}

We explore a spherically symmetric metric that satisfies the conformal killing vector, i.e., satisfying the equation
\begin{align}\label{41}
\mathcal{L}_{\chi} g_{\rho\nu}=2\varphi g_{\rho\nu}\rightarrow \mathcal{L}_{\chi}g^{\rho\nu}=-2\varphi g^{\rho\nu},
\end{align}
where $\varphi$ is a function of coordinates $t$ and $r$ and $\mathcal{L}_{\chi}$ symbolizes the Lie derivative with regard to the vector field $\chi$, whose general expression is
\begin{align*}
\chi=\zeta(t,r)\partial_{t}+\lambda(t,r)\partial_{r},
\end{align*}
where $\zeta$ and $\lambda$ are functions of independent variables $t$ and $r$. The case $\varphi=constant$ corresponds to a homothetic killing vector.
Our aim is to find exact solutions that admit a one-parameter group of conformal motions. We define it in terms of fundamental functions and initiate it by examining the case $\chi^{\upsilon}$ orthogonal to $V^{\upsilon}$ and $q=0$.

\subsection{Case of $\chi_{\upsilon}V^{\upsilon}=q=0$}

In this context, Eq. \eqref{41} provides
\begin{align}\label{42}
\mathcal{L_{\chi}}g_{\rho\nu}=2\varphi g_{\rho\nu}=\chi^{\alpha}\partial_{\alpha}g_{\rho\nu}+
g_{\rho\alpha}\partial_{\nu}\chi^{\alpha}+g_{\nu\alpha}\partial_{\rho}\chi^{\alpha},
\end{align}
which further implies
\begin{align}\label{43}
\varphi=\chi^{1}\frac{A'}{A},
\end{align}
\begin{align}\label{44}
\varphi=\chi^{1}\frac{B'}{B}+(\chi^{1})',
\end{align}
\begin{align}\label{45}
\varphi=\chi^{1}\frac{C'}{C},
\end{align}
and
\begin{align}\label{46}
\chi^{1}_{,t}=\chi^{1}_{,\theta}=\chi^{1}_{,\phi}=0.
\end{align}
From Eq. \eqref{43} and Eq. \eqref{45}, we obtain
\begin{align}\label{47}
A=n(t)C,
\end{align}
where $n$ is the integration function. We can take it $1$ if we reparameterize $t$. Therefore, we can write
\begin{align}\label{48}
A=\omega C,
\end{align}
where $\omega$ is the unit constant. The time derivatives of \eqref{44} and \eqref{45} along with \eqref{46} produce
\begin{align}\label{49}
\frac{B}{N(r)}=\omega N_{1}(t)C,
\end{align}
where $N(r)$ is arbitrary integration function which can be taken equal to $1$ and $ N_{1}(t)$
is an integration function that is dimensionless. Thus, we get
\begin{align}\label{50}
B=\omega N_{1}(t)C,
\end{align}
and
\begin{align}\label{51}
(\chi^{1})'=0\Rightarrow \chi^{1}=constant.
\end{align}
Using the above values in Eq. (A3) with $q=0$, we obtain
\begin{align}\label{52}
A=\omega C=\frac{F(t)}{k(t)+g(r)+I(r,t)},\quad\quad B=\frac{1}{k(t)+g(r)+I(r,t)},
\end{align}
where $k$, $g$ and $I$ are arbitrary functions of coordinates $t$ and $r$ and
\begin{align}\label{53}
F(t)\equiv\frac{1}{F_{1}(t)}.
\end{align}
Thus, any model can be determined by specifying the four arbitrary functions $F(t),~k(t),~g(r)$, and $I(r,t)$. The field equations in the form of arbitrary functions $F(t),~g(r),~k(t)$, and $I(r,t)$ can be written as
\begin{align}\nonumber
\mu=&\frac{f_{R}}{8\pi}\bigg[\frac{(k+g+I)^{2}}{F^{2}}\bigg\{\frac{\dot{F}^{2}}{F^{2}}-\frac{4(\dot{k}+\dot{I})}{k+g+I}+
\frac{3(\dot{k}+\dot{I})^{2}}{(k+g+I)^{2}}+\omega^{2}\bigg\}+
(k+g+I^{2})\bigg\{\frac{2(g''+I'')}{k+g+I}-\frac{3(g'+I')^{2}}{k+g+I}\bigg\}\bigg]
-\\\label{54}&\bigg[\frac{1}{8\pi}\bigg\{(k+g+I)^{2} f''_{R}-(k+g+I)(g'+I') f'_{R}-\frac{(k+g+I)^{2}}{F^{2}}
\bigg(\frac{2\dot{F}}{F}-\frac{3(\dot{k}+\dot{I})}{k+g+I}\bigg)\dot{f_{R}}-
\frac{f-Rf_{R}}{2}\bigg\}\bigg],\\\nonumber
P_{r}=&\frac{f_{R}}{8\pi}\bigg[-\frac{(k+g+I)^{2}}{F^{2}}\bigg\{\frac{2\ddot{F}}{F}-\frac{3\dot{F}^{2}}{F^{2}}
-\frac{2(\ddot{k}+\ddot{I})^{2}}{k+g+I}+\frac{3(\dot{k}+\dot{I})^{2}}{(k+g+I)^{2}}+\omega^{2}\bigg\}
+3(g'+I')^{2}\bigg]-\\\label{55}&\bigg[\frac{1}{8\pi}\bigg\{\frac{f-Rf_{R}}{2}+\frac{(k+g+I)^{2}}{F^{2}}
\bigg(\ddot{f_{R}}+\bigg(\frac{\dot{F}}{F}-\frac{(\dot{k}+\dot{I})}{k+g+I}\bigg)\dot{f_{R}}\bigg)
+3(g'+I')(k+g+I)f'_{R}\bigg\}\bigg],\\\nonumber
P_{\bot}=&\bigg[-\frac{(k+g+I)^{2}}{F^{2}}\bigg\{\frac{\ddot{F}}{F}-\frac{\dot{F}^{2}}{F^{2}}
-\frac{2(\ddot{k}+\ddot{I})^{2}}{k+g+I}+\frac{4(\dot{k}+\dot{I})^{2}}{(k+g+I)^{2}}\bigg\}+
(k+g+I)^{2}\bigg\{\frac{3(g'+I')^{2}}{(k+g+I)^{2}}-\frac{2(g''+I'')}{k+g+I}\bigg\}\bigg]-
\\\label{56}&\bigg[\frac{1}{8\pi}\bigg\{\frac{f-Rf_{R}}{2}+
\frac{(k+g+I)^{2}}{F^{2}}\ddot{f_{R}}-(k+g+I)^{2}f''_{R}
+(k+g+I)\frac{(\dot{k}+\dot{I})}{F^{2}}\dot{f_{R}}-(g'+I')(k+g+I)f'_{R}\bigg\}\bigg].
\end{align}
Imposing the junction conditions on the surface $r=r_{\Sigma}=constant$, we obtain from Eq.s \eqref{31} and \eqref{32}, respectively
\begin{align}\label{57}
\dot{C}^{2}_{\Sigma}+\omega^{2}(C^{2}_{\Sigma}-2MC_{\Sigma}-\alpha C^{4}_{\Sigma})=0,
\end{align}
and
\begin{align}\label{58}
2\ddot{C_{\Sigma}}C_{\Sigma}-\dot{C}^{2}_{\Sigma}-{\omega}^{2}(3\alpha C^{4}_{\Sigma}-C^{2}_{\Sigma})=0,
\end{align}
where
$\alpha\equiv g'^{2}(r_{\Sigma})$. Equation \eqref{57} can also be written as
\begin{align}\label{59}
\dot{C}^{2}_{\Sigma} =\omega^{2}C^{4}_{\Sigma}[\alpha-W(C_{\Sigma})],
\end{align}
where
\begin{align}\label{60}
W(C_{\Sigma})=\frac{1}{C^{2}_{\Sigma}}-\frac{2M}{C^{3}_{\Sigma}},
\end{align}
which can further be written as
\begin{align}\label{61}
M^{2}W(C_{\Sigma})=\frac{1}{z^{2}}-\frac{2}{z^{3}},
\end{align}
with
$z\equiv\frac{C_{\Sigma}}{M}$. Equation \eqref{59} can be manipulated as under
\begin{align}\label{62}
\omega(t-t_{0})=\pm\int\frac{dC_{\Sigma}}{C^{2}_{\Sigma}\sqrt{\alpha-W(C_{\Sigma})}}.
\end{align}
The integration of Eq. \eqref{62} gives the following two solutions (with $\alpha=\frac{1}{27M^{2}}$)
\begin{align}\label{63}
C_{\Sigma}^{(1)}=\frac{6M\tanh^{2}[\frac{\omega}{2}(t-t_{0})]}{3-\tanh^{2}[\frac{\omega}{2}(t-t_{0})]},\quad
C_{\Sigma}^{(2)}=\frac{6M\coth^{2}[\frac{\omega}{2}(t-t_{0})]}{3-\coth^{2}[\frac{\omega}{2}(t-t_{0})]}.
\end{align}
The first solution shows that the areal radius $C_{\Sigma}^{(1)}$ expands from $0$ to $3M$ as $t\rightarrow\infty$. It represents a white whole structure. However, in the second solution, the areal radius $C_{\Sigma}^{(2)}$ contracts from $\infty$ to $3M$ as $t\rightarrow\infty$. The entity $k(t)$ can be calculated from the expression $C_{\Sigma}$. While we impose quasi-homologous conditions and vanishing complexity factor conditions in order to determine other arbitrary functions. In the case $q=0$, the quasi-homologous condition implies $(\sigma=0)$, which further implies
\begin{align}\label{65}
\frac{\dot{B}}{B}=\frac{\dot{C}}{C}\Longrightarrow F(t)=\mathrm{Constant} \equiv F_{0}.
\end{align}
In this case, the metric functions take the form
\begin{align}\label{66}
A=\frac{F_{0}}{k(t)+g(r)+I(r,t)},\quad\quad B=\frac{1}{k(t)+g(r)+I(r,t)},\quad\quad   C=\frac{F_{0}}{\omega(k(t)+g(r)+I(r,t))}.
\end{align}
Now, instead of four arbitrary functions, we have to evaluate only three functions. One of these arbitrary functions can be fixed by imposing the vanishing complexity factor condition.
\begin{align}\label{67}
\frac{A''}{A}-\frac{A'}{A}\bigg(\frac{B'}{B}+\frac{C'}{C}\bigg)=0.
\end{align}
Using Eq. \eqref{66} in Eq. \eqref{67}, we obtain
\begin{align*}
g(r)+I(r,t)=c_{1}(t)r+c_{2}(t).
\end{align*}
From Eq. \eqref{63} and, we obtain
\begin{align}\label{68}
k^{(1)}(t)=\frac{F_{0}(3-\tanh^{2}[\frac{\omega}{2}(t-t_{0})])}{6\omega M\tanh^{2}[\frac{\omega}{2}(t-t_{0})]}-c_{1}(t)r_{\Sigma}-c_{2}(t),
\end{align}
and
\begin{align}\label{69}
k^{(2)}(t)=\frac{F_{0}(3-\coth{2}[\frac{\omega}{2}(t-t_{0})])}{6\omega M\coth^{2}[\frac{\omega}{2}(t-t_{0})]}-c_{1}(t)r_{\Sigma}-c_{2}(t).
\end{align}
The physical variables corresponding to $C^{(1)}_\Sigma$ are mentioned in Appendix B. The expansion scalar for this model is found as under
\begin{align}\label{74}
\Theta=\frac{3}{2M}\frac{\cosh[\frac{\omega}{2}(t-t_{0})]}{\sinh^{3}[\frac{\omega}{2}(t-t_{0})]}+3F_{0}\dot{c_{1}}
(t)(r_{\Sigma}-r),
\end{align}
thereby showing its homogenous and positive nature. The physical variables corresponding to $C^{(2)}_\Sigma$ are described in Appendix B. Now, we consider $\alpha=0$. For this case, the expression of $C_{\Sigma}$ takes the form
\begin{align}\label{78}
C^{(3)}_{\Sigma}=2M\cos^{2}[\frac{\omega}{2}(t-t_{0})],
\end{align}
also
\begin{align}\label{79}
\alpha=g'^{2}(r_{\Sigma})=c_{1}^{2}(t)=0.
\end{align}
Thus, the metric coefficients take the form
\begin{align}\label{80}
A=\frac{F_{0}}{k(t)+c_{2}(t)},\quad\quad    B=\frac{1}{k(t)+c_{2}(t)},\quad\quad   C=\frac{F_{0}}{\omega(k(t)+c_{2}(t))},
\end{align}
where $f(t)=f^{(3)}(t)$ is
\begin{align}\label{81}
k^{(3)}(t)=\frac{F_{0}}{2\omega M\cos^{2}[\frac{\omega}{2}(t-t_{0})]}-c_{2}(t),
\end{align}
The physical variables for this model are
\begin{align}\label{82}
&\mu=\frac{f_{R}}{8\pi}\bigg[\frac{3-2\cos^{2}[\frac{\omega}{2}(t-t_{0})]}{4M^{2}\cos^{6}[\frac{\omega}{2}
(t-t_{0})]}\bigg]-\bigg[\frac{1}{8\pi}\bigg\{\frac{F_{0}^{2}}{4\omega^{2}M^{2}cos^{4}
[\frac{\omega}{2}(t-t_{0})]}f''_{R}+\frac{F_{0}\sin[\frac{\omega}{2}(t-t_{0})]}{4\omega M^{2}\cos^{5}[\frac{\omega}{2}(t-t_{0})]}\dot{f_{R}}-\frac{f-Rf_{R}}{2}\bigg\}\bigg],\\\label{83}
&P_{r}=-\bigg[\frac{1}{8\pi}\bigg\{\frac{f-Rf_{R}}{2}-\frac{1}{4\omega^{2}M^{2}\cos^{4}[\frac{\omega}{2}(t-t_{0})]}\ddot{f_{R}}
-\frac{\sin[\frac{\omega}{2}(t-t_{0})]}{4\omega M^{2}\cos^{5}[\frac{\omega}{2}(t-t_{0})]}\dot{f_{R}}\bigg\}\bigg],\\\label{84}
&P_{\bot}=\frac{f_{R}}{8\pi}\bigg[\frac{1}{4M^{2}\cos^{4}[\frac{\omega}{2}(t-t_{0})]}\bigg]-\frac{1}{8\pi}\bigg[
\frac{f-Rf_{R}}{2}+\frac{1}{4\omega^{2}M^{2}\cos^{4}[\frac{\omega}{2}(t-t_{0})]}\bigg(\ddot{f_{R}}+F_{0}^{2}f''_{R}\bigg)
+\frac{\sin[\frac{\omega}{2}(t-t_{0})]}{4\omega M^{2}\cos^{5}[\frac{\omega}{2}(t-t_{0})]}\dot{f_{R}}\bigg].
\end{align}
Now we presume that the outside metric is Minkowski, implying that $M=0$. Thus, the solutions of Eq. \eqref{57} are as
\begin{align}\label{85}
C^{(4)}_{\Sigma}=\frac{1}{\sqrt{\alpha}\cos[\omega(t-t_{0})]},
\quad\quad
C^{(5)}_{\Sigma}=\frac{1}{\sqrt{\alpha}\sin[\omega(t-t_{0})]}.
\end{align}
From the above solutions, we obtain
\begin{align}\label{86}
k^{(4)}(t)=\frac{F_{0}}{\omega}\sqrt{\alpha}\cos[\omega(t-t_{0})]-c_{1}(t)r_{\Sigma}-c_{2}(t),
\quad\quad
k^{(5)}(t)=\frac{F_{0}}{\omega}\sqrt{\alpha}\sin[\omega(t-t_{0})]-c_{1}(t)r_{\Sigma}-c_{2}(t).
\end{align}
The physical variables corresponding to $C^{(4)}(t)$ read as
\begin{align}\nonumber
&\mu=\frac{f_{R}}{8\pi}\bigg[-2\alpha\cos^{2}[\omega(t-t_{0})]+c_{1}(t)(r_{\Sigma}-r)
\bigg[c_{1}(t)(r_{\Sigma}-r)-2\sqrt{\alpha}\cos[\omega(t-t_{0})]\bigg]+
\frac{3\dot{c_{1}}^{2}(t)(r_{\Sigma}-r)^{2}}{\omega^{2}r_{\Sigma}^{2}}+\\\nonumber&\frac{6}{\omega
r_{\Sigma}}\sqrt{\alpha}\dot{c_{1}}(t)(r_{\Sigma}-r)\sin[\omega(t-t_{0})]+3\alpha-3c_{1}^{2}(t)\bigg]-\frac{1}{8\pi}\bigg[
\bigg\{\frac{F_{0}\sqrt{\alpha}}{\omega}\cos[\omega(t-t_{0})]-c_{1}(t)(r_{\Sigma}-r)\bigg\}^{2}f''_{R}\\\nonumber
&-\bigg\{\frac{F_{0}\sqrt{\alpha}}{\omega}\cos[\omega(t-t_{0})]-c_{1}(t)(r_{\Sigma}-r)\bigg\}c_{1}(t)f'_{R}
+\frac{1}{F_{0}}\bigg\{\frac{F_{0}\sqrt{\alpha}}{\omega}\cos[\omega(t-t_{0})]-c_{1}(t)(r_{\Sigma}-r)\bigg\}
\\\label{87}
&\times\bigg\{-F_{0}\sqrt{\alpha}\sin[\omega(t-t_{0})]-\dot{c_{1}}(t)(r_{\Sigma}-r)\bigg\}\dot{f_{R}}-\frac{f-Rf_{R}}{2}\bigg],\\\nonumber
&P_{r}=\frac{f_{R}}{8\pi}\bigg[\frac{\omega}{F_{0}}(r_{\Sigma}-r)\bigg\{4\sqrt{\alpha}c_{1}(t)\cos[\omega(t-t_{0})]-
\frac{\omega}{F_{0}}c_{1}^{2}(t)(r_{\Sigma}-r)\bigg\}-\frac{2\sqrt{\alpha}}{\omega F_{0}}\cos[\omega(t-t_{0})]\ddot{c_{1}}(t)(r_{\Sigma}-r)
\\\nonumber
&+\frac{2}{F_{0}^{2}}c_{1}(t)\ddot{c_{1}}(t)(r_{\Sigma}-r)
^{2}-\frac{3}{F_{0}^{2}}\dot{c_{1}}^{2}(t)(r_{\Sigma}-r)^{2}-\frac{6}{F_{0}}\sqrt{\alpha}\dot{c_{1}}(t)(r_{\Sigma}-r)
\sin[\omega(t-t_{0})]+3\alpha+3c_{1}^{2}(t)\bigg]\\\nonumber&-\frac{1}{8\pi}\bigg[\frac{f-Rf_{R}}{2}-\frac{1}
{F_{0}^{2}}\bigg\{\frac{F_{0}\sqrt{\alpha}}{\omega}\cos[\omega(t-t_{0})]-c_{1}(t)(r_{\Sigma}-r)\bigg\}^{2}\ddot{f_{R}}
+\bigg\{\frac{1}{F_{0}^{2}}\bigg(\frac{F_{0}\sqrt{\alpha}}{\omega}\cos[\omega(t-t_{0})]\\\nonumber
&-c_{1}(t)(r_{\Sigma}-r)\bigg)
^{3}-\frac{F_{0}\sqrt{\alpha}\sin[\omega(t-t_{0})]-\dot{c_{1}(t)}(r_{\Sigma}-r)}{\frac{
F_{0}\sqrt{\alpha}}{\omega}\cos[\omega(t-t_{0})]-c_{1}(t)(r_{\Sigma}-r)}\bigg\}\dot{f_{R}}+3c_{1}(t)\bigg\{\frac
{F_{0}\sqrt{\alpha}}{\omega}\cos[\omega(t-t_{0})]-\\\label{88}&c_{1}(t)(r_{\Sigma}-r)\bigg\}f'_{R}\bigg],\\\nonumber
&P_{\bot}=\frac{f_{R}}{8\pi}\bigg[\alpha\cos^{2}[\omega(t-t_{0})]+\frac{2\omega\sqrt{\alpha}}{F_{0}}c_{1}(t)(r_{\Sigma}-r)
\cos[\omega(t-t_{0})]-\frac{2\sqrt{\alpha}}{\omega F_{0}}\ddot{c_{1}}(t)(r_{\Sigma}-r)\cos[\omega(t-t_{0})]\\\nonumber
&+\frac{2}{F_{0}^{2}}c_{1}(t)\ddot{c_{1}}(t)(r_{\Sigma}-r)^{2}-\frac{3}{F_{0}^{2}}\dot{c_{1}}^{2}(t)(r_{\Sigma}-r)^{2}
-3\alpha+3c_{1}^{2}(t)+\frac{6}{F_{0}}\sqrt{\alpha}\dot{c_{1}}(t)(r_{\Sigma}-r)\cos[\omega(t-t_{0})]\bigg]
\\\nonumber
&-\frac{1}{8\pi}\bigg[\frac{f-Rf_{R}}{2}+\bigg\{\frac{F_{0}\sqrt{\alpha}}{\omega}
\sin[\omega(t-t_{0})]-c_{1}(t)(r_{\Sigma}-r)\bigg\}^{2}\bigg\{\frac{1}{F_{0}^{2}}\ddot{f_{R}}+f''_{R}\bigg\}\\\nonumber&+\frac
{1}{F_{0}^{2}}\bigg\{\frac{F_{0}\sqrt{\alpha}}{\omega}\cos[\omega(t-t_{0})]-c_{1}(t)(r_{\Sigma}-r)\bigg\}
\bigg\{F_{0}\sqrt{\alpha}\cos[\omega(t-t_{0})]-\dot{c_{1}}(t)(r_{\Sigma}-r)\bigg\}\dot{f_{R}}\\\label{89}
&-
\bigg\{\frac{F_{0}\sqrt{\alpha}}{\omega}\cos[\omega(t-t_{0})]-c_{1}(t)(r_{\Sigma}-r)\bigg\}c_{1}(t)f_{R}\bigg].
\end{align}
The physical variables corresponding to $C^{(5)}(t)$ read as
\begin{align}\nonumber
&\mu=\frac{f_{R}}{8\pi}\bigg[-2\alpha\sin^{2}[\omega(t-t_{0})]+c_{1}(t)(r_{\Sigma}-r)
\bigg[c_{1}(t)(r_{\Sigma}-r)-2\sqrt{\alpha}\sin[\omega(t-t_{0})]\bigg]+
\frac{3\dot{c_{1}}^{2}(t)(r_{\Sigma}-r)^{2}}{\omega^{2}r_{\Sigma}^{2}}\\\nonumber
&+\frac{6}{\omega
r_{\Sigma}}\sqrt{\alpha}\dot{c_{1}}(t)(r_{\Sigma}-r)\cos[\omega(t-t_{0})]+3\alpha-3c_{1}^{2}(t)\bigg]-\frac{1}{8\pi}\bigg[
\bigg\{\frac{F_{0}\sqrt{\alpha}}{\omega}\sin[\omega(t-t_{0})]-c_{1}(t)(r_{\Sigma}-r)\bigg\}^{2}f''_{R}\\\nonumber
&-
\bigg\{\frac{F_{0}\sqrt{\alpha}}{\omega}\sin[\omega(t-t_{0})]-c_{1}(t)(r_{\Sigma}-r)\bigg\}c_{1}(t)f'_{R}
+\frac{1}{F_{0}}\bigg\{\frac{F_{0}\sqrt{\alpha}}{\omega}\sin[\omega(t-t_{0})]-c_{1}(t)(r_{\Sigma}-r)\bigg\}
\\\label{90}&
\bigg\{-F_{0}\sqrt{\alpha}\cos[\omega(t-t_{0})]-\dot{c_{1}}(t)(r_{\Sigma}-r)\bigg\}\dot{f_{R}}-\frac{f-Rf_{R}}{2}\bigg],\\\nonumber
&P_{r}=\frac{f_{R}}{8\pi}\bigg[\frac{\omega}{F_{0}}(r_{\Sigma}-r)\bigg\{4\sqrt{\alpha}c_{1}(t)\sin[\omega(t-t_{0})]-
\frac{\omega}{F_{0}}c_{1}^{2}(t)(r_{\Sigma}-r)\bigg\}-\frac{2\sqrt{\alpha}}{\omega F_{0}}\sin[\omega(t-t_{0})]\ddot{c_{1}}(t)(r_{\Sigma}-r)\\\nonumber
&+\frac{2}{F_{0}^{2}}c_{1}(t)\ddot{c_{1}}(t)(r_{\Sigma}-r)
^{2}-\frac{3}{F_{0}^{2}}\dot{c_{1}}^{2}(t)(r_{\Sigma}-r)^{2}-\frac{6}{F_{0}}\sqrt{\alpha}\dot{c_{1}}(t)(r_{\Sigma}-r)
\cos[\omega(t-t_{0})]+3\alpha+3c_{1}^{2}(t)\bigg]\\\nonumber
&-\frac{1}{8\pi}\bigg[\frac{f-Rf_{R}}{2}-\frac{1}
{F_{0}^{2}}\bigg\{\frac{F_{0}\sqrt{\alpha}}{\omega}\sin[\omega(t-t_{0})]-c_{1}(t)(r_{\Sigma}-r)\bigg\}^{2}\ddot{f_{R}}
+\bigg\{\frac{1}{F_{0}^{2}}\bigg(\frac{F_{0}\sqrt{\alpha}}{\omega}\sin[\omega(t-t_{0})]
\\\nonumber
&-c_{1}(t)(r_{\Sigma}-r)\bigg)
^{3}-\frac{F_{0}\sqrt{\alpha}\cos[\omega(t-t_{0})]-\dot{c_{1}(t)}(r_{\Sigma}-r)}{\frac{
F_{0}\sqrt{\alpha}}{\omega}\sin[\omega(t-t_{0})]-c_{1}(t)(r_{\Sigma}-r)}\bigg\}\dot{f_{R}}+3c_{1}(t)\bigg\{\frac
{F_{0}\sqrt{\alpha}}{\omega}\sin[\omega(t-t_{0})]-\\\label{91}&c_{1}(t)(r_{\Sigma}-r)\bigg\}f'_{R}\bigg],\\\nonumber
&P_{\bot}=\frac{f_{R}}{8\pi}\bigg[\alpha\sin^{2}[\omega(t-t_{0})]+\frac{2\omega\sqrt{\alpha}}{F_{0}}c_{1}(t)(r_{\Sigma}-r)
\sin[\omega(t-t_{0})]-\frac{2\sqrt{\alpha}}{\omega F_{0}}\ddot{c_{1}}(t)(r_{\Sigma}-r)\sin[\omega(t-t_{0})]\\\nonumber
&+
\frac{2}{F_{0}^{2}}c_{1}(t)\ddot{c_{1}}(t)(r_{\Sigma}-r)^{2}-\frac{3}{F_{0}^{2}}\dot{c_{1}}^{2}(t)(r_{\Sigma}-r)^{2}
-3\alpha+3c_{1}^{2}(t)+\frac{6}{F_{0}}\sqrt{\alpha}\dot{c_{1}}(t)(r_{\Sigma}-r)\sin[\omega(t-t_{0})]\bigg]\\\nonumber&
-\frac{1}{8\pi}\bigg[\frac{f-Rf_{R}}{2}+\bigg\{\frac{F_{0}\sqrt{\alpha}}{\omega}
\sin[\omega(t-t_{0})]-c_{1}(t)(r_{\Sigma}-r)\bigg\}^{2}\bigg\{\frac{1}{F_{0}^{2}}\ddot{f_{R}}+f''_{R}\bigg\}\\\nonumber
&+\frac
{1}{F_{0}^{2}}\bigg\{\frac{F_{0}\sqrt{\alpha}}{\omega}\cos[\omega(t-t_{0})]-c_{1}(t)(r_{\Sigma}-r)\bigg\}
\bigg\{F_{0}\sqrt{\alpha}\cos[\omega(t-t_{0})]-\dot{c_{1}}(t)(r_{\Sigma}-r)\bigg\}\dot{f_{R}}\\\label{92}
&-
\bigg\{\frac{F_{0}\sqrt{\alpha}}{\omega}\cos[\omega(t-t_{0})]-c_{1}(t)(r_{\Sigma}-r)\bigg\}c_{1}(t)f_{R}\bigg].
\end{align}
In model \textbf{4} and model \textbf{5}, we have used the relation $F_{0}=\omega r_{\Sigma}$. The symmetry assumed in subsection \textbf{A} lowers the metric variables (two functions of t, one function of r, and one function of t and r) to four functions. The matching requirements \eqref{35} and \eqref{36} are then reduced to a single differential equation \eqref{65}, the solution of which gives one of the four functions characterizing the metric. We made assumptions about the values of the equation's parameters to arrive at a solution defined in respect of fundamental functions. The remaining functions are determined by imposing other restrictions on the system. In these models, the extra dark source terms that come from theory are represented as the last terms on the right side of the equations wrapped in square brackets.

\subsection{Case of $\chi_{\upsilon}V^{\upsilon}=0;q\neq0$}

In this subsection, we investigate a dissipative case when $\chi_{\upsilon}V^{\upsilon}=0$. From Eq. \eqref{42}, we acquire
\begin{align}\label{93}
A=\omega C,
\end{align}
where $\omega$ is a unit constant whose dimension is $\frac{1}{length}$. And
\begin{align}\label{94}
F(t)B=\omega C,
\end{align}
where $F(t)$ is a function that appeared in the integration process and
\begin{align}\label{95}
{\chi}^{1}_{,1}=0\Rightarrow {\chi^{1}}=constant.
\end{align}
Using the above equations in \eqref{eq:A3} with $(q\neq0)$, we get
\begin{align}\label{96}
\frac{\dot{B}'}{B}-\frac{2\dot{B}B'}{B^{2}}=\frac{4\pi}{f_{R}}\bigg[qAB-\frac{1}{8\pi}\bigg(\dot{f'_{R}}
-\frac{A'}{A}\dot{f_{R}}-\frac{\dot{B}}{B}f'_{R}\bigg)\bigg].
\end{align}
The integration of the above equation produces
\begin{align}\label{97}
B=\frac{1}{k(t)+g(r)-4\pi\int\int I_{2}(r,t)drdt}=\frac{A}{F(t)}=\frac{\omega C}{F(t)},
\end{align}
where $k(t)$, $g(r)$ and $I_2(r,t)$ are functions of integration. To find an exact solution, we impose vanishing complexity factor condition $(Y_{TF}=0)$, which implies
\begin{align}\label{98}
\frac{1}{F^{2}}\bigg(\frac{\dot{F}\dot{B}}{FB}+\frac{\ddot{F}}{F}-\frac{\dot{F}^{2}}{F^{2}}\bigg)+
\frac{B''}{B}-2\bigg(\frac{B'}{B}\bigg)^{2}=0.
\end{align}
For simplification, we consider
\begin{align}\label{99}
\bigg(\frac{\dot{F}\dot{B}}{FB}+\frac{\ddot{F}}{F}-\frac{\dot{F}^{2}}{F^{2}}\bigg)=0,
\end{align}
and
\begin{align}\label{100}
\frac{B''}{B}-2\bigg(\frac{B'}{B}\bigg)^{2}=0.
\end{align}
The integration of Eq. \eqref{100} leads to
\begin{align}\label{101}
B=-\frac{1}{\delta(t)r+\eta(t)},
\end{align}
where $\delta(t)$ and $\eta(t)$ are arbitrary integration functions. The dimension of $\delta(t)$ is $\frac{1}{length}$ and $\eta(t)$ is dimensionless. The differentiation of Eq. \eqref{99} gives $\eta=\frac{\delta}{\omega}$. Therefore, we can write Eq. \eqref{101} as
\begin{align}\label{102}
B=-\frac{\omega}{\delta(t)+(\omega r+1)}.
\end{align}
From Eqs. \eqref{99} and \eqref{102}, we obtain
\begin{align}\label{103}
F(t)=c_{3}e^{c_{4}\int\delta(t)dt},
\end{align}
where $c_{3}$ and $c_{4}$ are constants of integration. From the above Eq. \eqref{103}, we can write
\begin{align}\label{104}
\frac{\dot{F}}{F}=c_{4}\delta(t).
\end{align}
The comparison of Eqs. \eqref{97} and \eqref{102} gives
\begin{align}\label{105}
4\pi\int\int I_{2}drdt=\eta(t)\delta(t),\quad\quad    \delta(t)=-\omega k(t),
\end{align}
In the form of arbitrary function $\delta(t)$, the field equations (A1)-(A4) take the form as
\begin{align}\nonumber
\mu=&\frac{f_{R}}{8\pi}\bigg[\frac{\delta^{2}(\omega r+1)^{2}}{\omega^{2}c_{3}^{2}e^{2c_{4}\int\delta(t)dt}}
\bigg(\omega^{2}-4c_{4}\dot{\delta}+\frac{3\dot{\delta^{2}}}{\delta^{2}}+c_{4}^{2}\delta^{2}\bigg)
-3\delta^{2}\bigg]\\\label{106}&-\frac{1}{8\pi}\bigg[\frac{\delta^{2}(\omega r+1)^{2}}{\omega^{2}}f''_{R}
+\frac{\delta^{2}(\omega r+1)^{2}}{\omega}f'_{R}
-\frac{\delta^{2}(\omega r+1)^{2}}{\omega^{2}c_{3}^{2}e^{2c_{4}\int\delta(t)dt}}
\bigg(2c_{4}\delta-3\frac{\dot{\delta}}{\delta}\bigg)\dot{f_{R}}-\frac{f-Rf_{R}}{2}\bigg],\\\label{107}
q=&\frac{f_{R}}{4\pi}\bigg(-\frac{\delta\dot{\delta}(\omega r+1)}{\omega c_{3}e^{c_{4}\int\delta(t)dt}}\bigg)+
\frac{\delta^{2}(\omega r+1)^{2}}{8\pi\omega^{2}c_{3}e^{c_{4}\int\delta(t)dt}}
\bigg\{\dot{f'_{R}}+\frac{\omega}{\omega r+1}\dot{f_{R}}+\frac{\dot{\delta}}{\delta}f'_{R}\bigg\},\\\nonumber
P_{r}=&\frac{f_{R}}{8\pi}\bigg[-\frac{\delta^{2}(\omega r+1)^{2}}{\omega^{2}c_{3}^{2}e^{2c_{4}\int\delta(t)dt}}\bigg(\omega^{2}- \frac{2\ddot{\delta}}{\delta}+\frac{3\dot{\delta^{2}}}{\delta^{2}}+c_{4}^{2}\delta^{2}\bigg)+3\delta^{2}\bigg]
\\\label{108}&-\frac{1}{8\pi}\bigg[\frac{f-Rf_{R}}{2}-
\frac{\delta^{2}(\omega r+1)^{2}}{\omega^{2}c_{3}e^{c_{4}\int\delta(t)dt}}
\bigg\{\ddot{f_{R}}-\bigg(c_{4}\delta-\frac{\dot{\delta}}{\delta}\bigg)\dot{f_{R}}\bigg\}+\frac{3\delta^{2}(\omega r+1)}{\omega}f_{R}'\bigg],\\\nonumber
P_{\bot}=&\frac{f_{R}}{8\pi}\bigg[-\frac{\delta^{2}(\omega r+1)^{2}}{\omega^{2}c_{3}^{2}e^{2c_{4}\int\delta(t)dt}}\bigg(c_{4}\dot{\delta}- \frac{2\ddot{\delta}}{\delta}+\frac{3\dot{\delta^{2}}}{\delta^{2}}\bigg)+3\delta^{2}\bigg] \\\label{109}&-\frac{1}{8\pi}\bigg[\frac{f-Rf_{R}}{2}+\frac{\delta^{2}(\omega r+1)^{2}}
{\omega^{2}c_{3}e^{c_{4}\int\delta(t)dt}}\bigg(\ddot{f_{R}}-\frac{\dot{\delta}}{\delta}\dot{f_{R}}
\bigg)-\frac{\delta^{2}(\omega r+1)^{2}}{\omega^{2}}\bigg(f''_{R}+\frac{\omega}{\omega r+1}f'_{R}\bigg)\bigg].
\end{align}
The arbitrary function $\delta(t)$ can be found by imposing the quasi-homologous condition \eqref{43} and the junction condition \eqref{32}. By utilizing the quasi-homologous condition \eqref{40}, we get
\begin{align}\label{110}
\frac{\dot{F}}{F}=\frac{4\pi AB}{B'f_{R}}\bigg[qB-\frac{1}{8\pi A}\bigg(\dot{f'_{R}}-\frac{B'}{B}\dot{f_{R}}
-\frac{\dot{B}}{B}f'_{R}\bigg)\bigg].
\end{align}
Thus, the function $\delta(t)$ can be written as
\begin{align}\label{111}
\delta=\frac{1}{c_{4}t+c_{5}},
\end{align}
where $c_{5}$ is an integration constant whose dimension is length. Resultantly, the function $B$ is formulated as
\begin{align}\label{112}
B=\frac{\omega(c_{4}t+c_{5})}{\omega r+1},
\end{align}
To solve a set of equations we consider the shear-free case under which $\sigma=0$. Now, the shear-free condition $(\sigma=0)$ implies that $\dot{F}=0$ which further implies that $c_{4}=0$. Thus, the metric coefficients are calculated as
\begin{align}\label{113}
A=c_{3}B=\frac{-\omega c_{3}}{\delta(\omega r+1)},\quad\quad C=\frac{-c_{3}}{\delta(\omega r+1)},
\end{align}
The physical variables in the form of the arbitrary function $\delta(t)$ take the form
\begin{align}\label{114}
&\mu=\frac{f_{R}}{8\pi}\bigg[\frac{\delta^{2}(\omega r+1)^{2}}{\omega^{2}c_{3}^{2}}\bigg(\omega^{2}+
\frac{3\dot{\delta^{2}}}{\delta^{2}}\bigg)-3\delta^{2}\bigg]-\frac{1}{8\pi}\bigg[\frac{\delta^{2}
(\omega r+1)^{2}}{\omega^{2}}\bigg(f''_{R}-\frac{\omega}{\omega r+1}f'_{R}+\frac{3\dot{\delta}}{c_{3}^{2}\delta}\dot{f_{R}}\bigg)-\frac{f-Rf_{R}}{2}\bigg],\\\label{115}
&q=\frac{f_{R}}{8\pi}\frac{-\delta\dot{\delta}(\omega r+1)}{\omega c_{3}}+\frac{1}{8\pi}\bigg[\frac{\delta^{2}(\omega r+1)^{2}}{2c_{3}\omega^{2}}\bigg(\dot{f'_{R}}+\frac{\omega}{\omega r+1}\dot{f_{R}}+\frac{\dot{\delta}}{\delta}f'_{R}\bigg)\bigg],\\\nonumber
&P_{r}=\frac{f_{R}}{8\pi}\bigg[\frac{-\delta^{2}(\omega r+1)^{2}}{\omega^{2}c_{3}^{2}}
\bigg(\omega^{2}-\frac{2\ddot{\delta}}{\delta}+\frac{3\dot{\delta^{2}}}{\delta^{2}}\bigg)+3\delta^{2}\bigg]
-\frac{1}{8\pi}\bigg[\frac{f-Rf_{R}}{2}+\frac{\delta^{2}(\omega r+1)^{2}}{c_{3}^{2}\omega^{2}}\\\label{116}&\times
\bigg(\ddot{f_{R}}-\frac{\dot{\dot{\delta}}}{\delta}\dot{f_{R}}+3c_{3}^{2}\frac{\omega}{\omega r+1}f'_{R}\bigg)\bigg],\\
\nonumber
&P_{\bot}=\frac{f_{R}}{8\pi}\bigg[\frac{-\delta^{2}(\omega r+1)^{2}}{\omega^{2}c_{3}^{2}}
\bigg(\frac{-2\ddot{\delta}}{\delta}+\frac{3\dot{\delta^{2}}}{\delta^{2}}\bigg)+3\delta^{2}\bigg]-
\frac{1}{8\pi}\bigg[\frac{f-Rf_{R}}{2}+\frac{\delta^{2}(\omega r+1)^{2}}{c_{3}^{2}\omega^{2}}
\bigg(\ddot{f_{R}}+\frac{\dot{\delta}}{\delta}f_{R}\bigg)+\\\label{117}&\frac{\delta^{2}(\omega r+1)^{2}}{\omega^{2}}\bigg(f''_{R}+\frac{\omega}{\omega r+1}f'_{R}\bigg)\bigg].
\end{align}
To specify the function $\delta(t)$, we impose junction condition \eqref{33}. Using the Eqs. \eqref{114}-\eqref{117} in Eq. \eqref{32}, we obtain
\begin{align}\label{118}
2\frac{\ddot{\delta}}{\delta}-3\bigg(\frac{\dot{\delta}}{\delta}\bigg)^{2}+\frac{2\omega_{1}\dot{\delta}}{\delta}=
\omega^{2}-3\omega_{1}^{2},
\end{align}
with $\omega_{1}\equiv\frac{\omega c_{3}}{\omega r_{\Sigma}+1}$. To find the solution of Eq. \eqref{118}, let $u=\frac{\dot{\delta}}{\delta}$. In this case, the above Eq. \eqref{118} becomes Ricatti equation, formulated as
\begin{align}\label{119}
2\dot{u}-u^{2}+2\omega_{1}u=\omega^{2}-3\omega_{1}^{2}.
\end{align}
The solution of this equation is found as under
\begin{align}\label{120}
u=\omega_{1}+\sqrt{\omega^{2}-4\omega_{1}^{2}}\tan\bigg[\frac{\sqrt{\omega^{2}-4\omega_{1}^{2}}}{2}
(t-t_{0})],
\end{align}
from which we can find $\delta$ as
\begin{align}\label{121}
\delta(t)=\omega_{2}e^{\omega_{1}t} \sec^{2}\bigg[\frac{\sqrt{\omega^{2}-4\omega_{1}^{2}}}{2}(t-t_{0})],
\end{align}
where $\omega_{2}$ an is an integration constant, whose dimension is the same as of $\delta$. The expression of temperature is found by utilizing the transport Eq. \eqref{38} as
\begin{align}\label{122}
\mathcal{T}(r,t)=\frac{\omega r+1}{4\pi c_{3} \mathcal{K} \omega}\int\bigg[\frac{\tau(\dot{\delta^{2}}+\delta\ddot{\delta})}{c_{3}}-
\frac{\dot{\delta}\omega}{\omega r+1}\bigg]f_{R} dr-\frac{\delta(\omega r+1)}{\omega c_{3}}\int Z_{1}(r,t)dr+\mathcal{T}_{0}(t),
\end{align}
where $c_{3}$ and $\mathcal{T}_{0}(t)$ are arbitrary entities appeared due to integration process. In this instance, we discovered a model that evolves quasi-homologously \eqref{40}, the diminishing complexity factor $(Y_{TF}=0)$, and the matching conditions \eqref{31} and \eqref{32}, which together formulate all of the metric coefficients. Expressions \eqref{114}-\eqref{117} and expression \eqref{122}, which calculates temperature using the truncated transport equation \eqref{38}, define this model (model \textbf{6}). The additional dark source terms resulting from the theory are shown as the last terms on the right side of the equations enclosed in square brackets in this model. This model does not meet the regularity requirements at the center.

\subsection{Case of $\chi^{\upsilon}\|V^{\upsilon};q=0$}

In this subsection, we study the non-dissipative case with $\chi^{\upsilon}\|V^{\upsilon}$. In this respect, Eq. \eqref{42} gives
\begin{align}\label{123}
A=Bl(r),\quad\quad C=Br,\quad\quad \chi^{0}=1,\quad\quad\psi=\frac{\dot{B}}{B},
\end{align}
where $l(r)$ is an integration function. Resultantly, we can write the line element, mentioned in Eq. \eqref{6}, as
\begin{align}\label{124}
 ds^{2}=B^{2}[-l^{2}(r)dt^{2}+dr^{2}+r^{2}(d\theta^{2}+\sin^{2}\theta d\varphi^{2})].
\end{align}
Again utilizing \eqref{123}, Eq. \eqref{eq:A3} takes the form
\begin{align}\label{125}
\frac{\dot{C'}}{C}-2\frac{\dot{C}}{C}\frac{C'}{C}-\frac{\dot{C}}{C}\bigg(\frac{l'}{l}-\frac{1}{r})=0,
\end{align}
whose solution is found as under
\begin{align}\label{126}
C=\frac{r}{l(r)[k(t)+g(r)+I_{2}(r,t)]},
\end{align}
which implies
\begin{align}\label{127}
B=\frac{1}{l(r)[k(t)+g(r)+I_{2}(r,t)]},\quad\quad
A=\frac{1}{k(t)+g(r)+I_{2}(r,t)},
\end{align}
here $g(r)$ and $k(t)$ are arbitrary integration functions. To find the solution, we have to specify four arbitrary functions $k(t)$, $g(r)$, $l(r)$, and $I_{2}(r,t)$. The function $k(t)$ shall be evaluated by using the matching conditions Eq. \eqref{31} and Eq. \eqref{32}. The first condition reads
\begin{align}\label{128}
\dot{C^{2}_{\Sigma}}=\omega^{2}C^{4}_{\Sigma}[\varepsilon-W(C_{\Sigma})],
\end{align}
where $\omega^{2}\equiv\frac{l^{2}_{\Sigma}}{r^{2}_{\Sigma}}$, $\varepsilon\equiv(g')^{2}_{\Sigma} l^{2}_{\Sigma}$ , and
\begin{align}\label{129}
W(C_{\Sigma})=\frac{2\sqrt{\varepsilon}}{C_{\Sigma}}(1-a_{1})+\frac{a_{1}}{C_{\Sigma}^{2}}(2-a_{1})-\frac{2M}
{C_{\Sigma}^{3}},
\end{align}
with $a_{1}\equiv\frac{l'_{\Sigma}r_{\Sigma}}{l_{\Sigma}}.$
\begin{align}\label{130}
2\ddot{C_{\Sigma}}C_{\Sigma}-\dot{C^{2}_{\Sigma}}-3\varepsilon\omega^{2}C^{4}_{\Sigma}-
4\omega^{2}\sqrt{\varepsilon}
C^{3}_{\Sigma}(a_{1}-1)-\omega^{2}C^{2}_{\Sigma}a_{1}(a_{1}-2)=0.
\end{align}
Now, we have to find the solution of Eq. \eqref{130} in order to specify the arbitrary function $k(t)$. We consider a particular case $a_{1}=0$, which implies
\begin{align}\label{131}
\dot{C^{2}_{\Sigma}}=\omega^{2}C^{4}_{\Sigma}\bigg[\varepsilon-\frac{1}{C^{2}_{\Sigma}}+
\frac{2M}{C^{3}_{\Sigma}}\bigg],
\end{align}
which is same as Eq. \eqref{59}, so have the same solutions.
The Eq. \eqref{32} produces
\begin{align}\label{132}
2\ddot{C_{\Sigma}}C_{\Sigma}-\dot{C^{2}_{\Sigma}}-3\varepsilon\omega^{2}C^{4}_{\Sigma}
+\omega^{2}C^{2}_{\Sigma}=0,
\end{align}
here Eq. \eqref{132} is just the derivative of Eq. \eqref{131}. Thus, we will just consider Eq. \eqref{132} for further calculation. Now, we will imply vanishing complexity factor condition \eqref{29} to specify the other arbitrary functions $g(r)$ and $l(r)$, which implies
\begin{align}\label{133}
\frac{2l'}{l}+\frac{g''+I_{2}''}{g'+I'}-\frac{1}{r}=0,
\end{align}
or
\begin{align}\label{134}
\frac{2l'}{l}+\frac{u'}{u}-\frac{1}{r}=0,
\end{align}
where $u\equiv g'+I'$. The solution of Eq. \eqref{134} is
\begin{align}\label{135}
u=\frac{b_{4}(t)r}{l^{2}},
\end{align}
giving us $g+I_{2}=b_{4}(t)\int\frac{r}{l^{2}}dr+b_{5}(t)$, with $b_{4}(t)$ and $ b_{5}(t)$ as integration functions. On selecting $l(r)=b_{6}r$, we get $a_{1}=1$ and  $g+I_{2}=b_{7}(t) \ln(r)+b_{5}(t)$, where $b_{6}$ and $b_{7}(t)$ are integration entities. The function $k(t)$ can be determined by imposing junction condition \eqref{31}, whose solution is given by
\begin{align}\label{136}
C^{(7)}_{\Sigma}=\frac{6M\tanh^{2}[\frac{\omega}{2}(t-t_{0})]}{3-\tanh^{2}[\frac{\omega}{2}(t-t_{0})]}.
\end{align}
The physical variables that correspond to this model are
\begin{align}\nonumber
&\mu=\frac{f_{R}}{8\pi}\bigg[3\bigg\{\frac{\coth[\frac{\omega}{2}(t-t_{0})]}{2M\sinh^{2}
[\frac{\omega}{2}(t-t_{0})]}+
\dot{b_{7}(t)}\ln\bigg(\frac{r}{r_{\Sigma}}\bigg)\bigg\}^{2}+b_{6}^{2}\bigg\{\frac{3\coth^{2}
[\frac{\omega}{2}(t-t_{0})]-1}{6b_{6}M}+
b_{7}(t)\ln\bigg(\frac{r}{r_{\Sigma}}\bigg)\bigg\}^{2}-3b_{6}^{2}b_{7}^{2}(t)\bigg]\\\nonumber
&-\frac{1}{8\pi}\bigg[b_{6}^{2}r^{2}
\bigg\{\frac{3\coth^{2}[\frac{\omega}{2}(t-t_{0})]-1}{6b_{6}M}+c_7(t)\ln\bigg(\frac{r}{r_{\Sigma}}\bigg)\bigg\}
\bigg\{f''_{R}+\bigg(\frac{1}{r}-\\\nonumber&\frac{6b_{6}b_{7}(t)M}{r\bigg\{3\coth^{2}[\frac{\omega}{2}(t-t_{0})]-1+
6b_{6}b_{7}(t)M\ln\bigg(\frac{r}{r_{\Sigma}}\bigg)\bigg\}}\bigg)f'_{R}\bigg\}+3\bigg\{\frac{-\omega
\coth[\frac{\omega}{2}(t-t_{0})]}{2b_{6}M\sinh^{2}[\frac{\omega}{2}(t-t_{0})]}+\dot{b}_7(t)\ln\bigg(\frac{r}
{r_{\Sigma}}\bigg)
\bigg\}\\\label{137}&\bigg\{\frac{3\coth^{2}[\frac{\omega}{2}(t-t_{0})]-1}{6b_{6}M}+b_{7}(t)\ln\bigg(\frac{r}{r_{\Sigma}}
\bigg)\bigg\}\dot{f_{R}}
-\frac{f-Rf_{R}}{2}\bigg],\\\nonumber
&P_{r}=\frac{f_{R}}{8\pi}\bigg[b_{6}b_{7}(t)\ln\bigg(\frac{r}{r_{\Sigma}}\bigg)\bigg[\frac{(3\coth^{2}
[\frac{\omega}{2}(t-t_{0})]-1)(3\coth^{2}[\frac{\omega}{2}
(t-t_{0})]-5)}{6M}-b_{6}b_{7}(t)\ln\bigg(\frac{r}{r_{\Sigma}}\bigg)\bigg]-\frac{1}{9M^{2}}\\\nonumber&
+3b^{2}_{6}b^{2}_{7}(t)+\frac{\dot{b}_{7}}{M}\ln\bigg(\frac{r}{r_{\Sigma}}\bigg)\frac{\coth^[\frac{\omega}
{2}(t-t_{0})]}{\sinh^{2}[\frac{\omega}{2}(t-t_{0})]}+2\ddot{b}_{7}(t)\ln\bigg(\frac{r}{r_{\Sigma}}\bigg)
\bigg\{\frac{(3\coth^{2}[\frac{\omega}{2}(t-t_{0})]-1)}{6b_{6}M}+b_{7}(t)\ln\bigg(\frac{r}{r_{\Sigma}}\bigg)
\bigg\}\bigg]\\\nonumber&
-\frac{1}{8\pi}\bigg[\frac{f-Rf_{R}}{2}+\bigg\{\frac{3\coth^{2}[\frac{\omega}{2}(t-t_{0})]-1}{6b_{6}M}
+b_7(t)\ln\bigg(\frac{r}{r_{\Sigma}}\bigg)\bigg\}^{2}\bigg\{\ddot{f_{R}}-\frac{\bigg\{\frac{-\omega
\coth[\frac{\omega}{2}(t-t_{0})]}{2b_{6}M\sinh^{2}[\frac{\omega}{2}(t-t_{0})]}+\dot{b}_7(t)\ln\bigg(\frac{r}
{r_{\Sigma}}\bigg)\bigg\}}{\bigg\{\frac{3\coth^{2}[\frac{\omega}{2}(t-t_{0})]-1}{6b_{6}M}
+b_7(t)\ln\bigg(\frac{r}{r_{\Sigma}}\bigg)\bigg\}}\\\label{138}&\dot{f}_{R}\bigg\}+3rb^{2}_{6}b^{2}_{7}(t)\bigg\{\frac{3\coth^{2}[\frac{\omega}{2}(t-t_{0})]-1}{6b_{6}M}
+b_7(t)\ln\bigg(\frac{r}{r_{\Sigma}}\bigg)\bigg\}f'_{R}\bigg],\\\nonumber
&P_{\bot}=\frac{f_{R}}{8\pi}\bigg[\frac{(3\coth^{2}[\frac{\omega}{2}(t-t_{0})]+1)}{12M^{2}\sinh^{2}[\frac
{\omega}{2}(t-t_{0})]}+\ln\bigg(\frac{r}{r_{\Sigma}}\bigg)\frac{(3\coth^{2}[\frac{\omega}{2}
(t-t_{0})]+1)}{6\sqrt{3}M^{2}\sinh^{2}[\frac{\omega}{2}(t-t_{0})]}+3b^{2}_{6}b^{2}_{7}(t)\\\nonumber&+\frac{3
\dot{b}_{7}}{M}\ln\bigg(\frac{r}{r_{\Sigma}}\bigg)\bigg\{\frac{\coth^[\frac{\omega}{2}(t-t_{0})]}
{\sinh^{2}[\frac{\omega}{2}(t-t_{0})]}-\dot{b}_{7}M\ln\bigg(\frac{r}{r_{\Sigma}}\bigg)\bigg\}
+2\ddot{b}_{7}(t)\ln\bigg(\frac{r}{r_{\Sigma}}\bigg)
\bigg\{\frac{(3\coth^{2}[\frac{\omega}{2}(t-t_{0})]-1)}{6b_{6}M}+\\\nonumber&b_{7}(t)\ln\bigg(\frac{r}
{r_{\Sigma}}\bigg)\bigg\}\bigg]-\frac{1}{8\pi}\bigg[\frac{f-Rf_{R}}{2}+\bigg\{\frac{3\coth^{2}[\frac
{\omega}{2}(t-t_{0})]-1}{6b_{6}M}+b_7(t)\ln\bigg(\frac{r}{r_{\Sigma}}\bigg)\bigg\}^{2}\bigg(
\ddot{f_{R}}-r^{2}b^{2}_{6}f''_R\bigg)\\\nonumber&-\bigg\{\frac{3\coth^{2}[\frac{\omega}{2}(t-t_{0})]-1}{6b_{6}M}+
b_7(t)\ln\bigg(\frac{r}{r_{\Sigma}}\bigg)\bigg\}\bigg\{\frac{-\omega\coth[\frac{\omega}{2}(t-t_{0})]}
{2b_{6}M\sinh^{2}[\frac{\omega}{2}(t-t_{0})]}+\dot{b}_7(t)\ln\bigg(\frac{r}
{r_{\Sigma}}\bigg)\bigg\}\dot{f}_R\\\label{139}&-rb^{2}_{6}(1+b_{7}(t))\bigg\{\frac{3\coth^{2}[\frac
{\omega}{2}(t-t_{0})]-1}{6b_{6}M}+b_7(t)\ln\bigg(\frac{r}{r_{\Sigma}}\bigg)\bigg\}f'_{R}\bigg].
\end{align}
Here, we have used the relation $\omega^{2}=c^{2}_{6}$. Now, we consider another case $\varepsilon=0$ and $a_{1}=\frac{1}{2}$, which implies
\begin{align}\label{140}
\dot{C^{2}_{\Sigma}}=\omega^{2}C^{4}_{\Sigma}\bigg[\frac{2M}{C^{3}_{\Sigma}}-
\frac{3}{4C^{2}_{\Sigma}}\bigg].
\end{align}
The integration of Eq. \eqref{140} gives
\begin{align}\label{141}
C^{(8)}_{\Sigma}=\frac{4M}{3}(1+sin T),
\end{align}
where $T\equiv\frac{\sqrt{3}\omega}{2}(t-t_{0})$. To determine other arbitrary functions, we shall vanish the complexity factor \eqref{33}. For $a_{1}=\frac{1}{2}$, we obtain
\begin{align}\label{142}
l(r)=b_{1}\sqrt{r},\quad\quad g(r)+I_{2}(r,t)=b_2(t)r+b_3(t).
\end{align}
Since $\varepsilon=0 \Rightarrow b_2(t)=0$, therefore $g(r)+I_{2}(r,t)=b_3(t)$. For this case, the physical variables are
\begin{align}\nonumber
&\mu=\frac{f_{R}}{8\pi}\bigg[\frac{27}{64M^{2}(1+\sin T)^{2}}\bigg\{\frac{3\cos^{2}T}{(1+\sin T)^{2}}+
\frac{r_{\Sigma}}{r}\bigg\}\bigg]-\frac{1}{8\pi}\bigg[\frac{9r r_{\Sigma}}{16M^{2}(1+\sin T)^{2}}\bigg\{f''_{R}-\frac{3}{2r}f'_{R}\bigg\}
+3b_2(t)\\\label{143}&
\times\bigg(\frac{3\sqrt{r_{\Sigma}}}{4Mb_{1}(1+sin T)}\bigg)\dot{f_{R}}
-\frac{f-Rf_{R}}{2}\bigg],\\\nonumber
&P_{r}=\frac{f_{R}}{8\pi}\frac{27}{64M^{2}(1+\sin T)^{2}}\bigg[1-\frac{r_{\Sigma}}{r}\bigg]-
 \frac{1}{8\pi}\bigg[\frac{f-Rf_{R}}{2}+\frac{9r_{\Sigma}}{4M^{2}b_{1}^{2}(1+\sin T)^{2}}\bigg\{\ddot{f_{R}}+\frac{\sqrt{3}l^{2}(r_{\Sigma})\cos T}{2r_{\Sigma}^{2}(1+\sin T)^{2}}\dot{f_{R}}\\\label{144}&-\frac{9r_{\Sigma}}{32M^{2}(1+\sin T)^{2}}f'_{R}\bigg\}\bigg],\\\nonumber
&P_\perp=\frac{f_{R}}{8\pi}\frac{27}{64M^{2}(1+\sin T)^{2}}-\frac{1}{8\pi}\bigg[\frac{f-Rf_{R}}{2}+
\frac{9r_{\Sigma}}{16M^{2}b_{1}^{2}(1+\sin T)^{4}}\bigg(\ddot{f_{R}}+rb_{1}^{2}f''_{R}\bigg)+\frac{9\sqrt{3r_{\Sigma}}\cos T}{32M^{2}b_{1}(1+\sin T)^{3}}\dot{f_{R}}\\\label{145}&
-\frac{9r_{\Sigma}}{16M^{2}(1+\sin T)^{4}}f'_{R}\bigg].
\end{align}
Now, we examine the case $M=0$, then Eq. \eqref{128} becomes
\begin{align}\label{146}
\dot{C^{2}_\Sigma}=\omega^{2}C^{4}_\Sigma\bigg[\varepsilon-\frac{2\sqrt{\varepsilon}(1-a_{1})}{C_\Sigma}-\frac{a_{1}(2-a_{1})}
{C^{2}_\Sigma}\bigg].
\end{align}
By considering the case $a_{1}=0$, the two solutions of above Eq. in the form of elementary functions are expressed as
\begin{align}\label{147}
C^{(9)}_\Sigma=\frac{1}{\sqrt{\varepsilon}\cos[\omega(t-t_{0})]},\quad\quad
C^{(10)}_\Sigma=\frac{1}{\sqrt{\varepsilon}\sin[\omega(t-t_{0})]}.
\end{align}
The physical variables corresponding to $C^{(9)}_\Sigma$ are calculated as
\begin{align}\nonumber
&\mu=\frac{f_{R}}{8\pi}\bigg[-2\varepsilon\cos^{2}[\omega(t-t_{0})]+b_{6}b_{7}(t)\ln\bigg(\frac{r}{r_{\Sigma}}\bigg)
\bigg\{b_{6}b_{7}(t)\ln\bigg(\frac{r}{r_{\Sigma}}\bigg)+2\sqrt{\varepsilon}\cos[\omega(t-t_{0})]\bigg\}
+3\varepsilon-3b_{6}^{2}b_{7}^{2}(t)\\\nonumber&
+3{\dot{c_7}^{2}(t)}\bigg\{\ln\bigg(\frac{r}{r_{\Sigma}}\bigg)\bigg\}^{2}
-6\sqrt{\varepsilon}\dot{b_{7}}(t)\sin[\omega(t-t_{0})]\ln\bigg(\frac{r}{r_{\Sigma}}\bigg)\bigg]-\frac{1}{8\pi}
\bigg[\bigg\{r^{2}\bigg(\sqrt{\varepsilon}\cos[\omega(t-t_{0})]+b_{6}b_{7}(t)
\\\nonumber&\times\ln\bigg(\frac{r}
{r_{\Sigma}}\bigg)\bigg)^{2}\bigg\}\bigg\{f''_{R}+\bigg(\frac{1}{r}-\frac{b_{6}b_{7}(t)}{r\bigg\{\sqrt{\varepsilon}
\cos[\omega(t-t_{0})]+b_{6}b_{7}(t)\ln\bigg(\frac{r}{r_{\Sigma}}\bigg)\bigg\}}\bigg)f'_{R}\bigg\}
\frac{3}{b_{6}}\bigg\{-\sqrt{\varepsilon}\sin[\omega(t-t_{0})]\\\label{148}&+\dot{b_{7}}(t)\ln\bigg(\frac{r}{r_{\Sigma}}\bigg)\bigg\}
\bigg\{\sqrt{\varepsilon}\cos[\omega(t-t_{0})]+b_{6}b_{7}(t)\ln\bigg(\frac{r}{r_{\Sigma}}\bigg)\bigg\}\dot{f_{R}}-
\frac{f-Rf_{R}}{2}\bigg],
\end{align}
\begin{align}\nonumber
P_{r}=&\frac{f_{R}}{8\pi}\bigg[-b_{6}b_{7}(t)\ln\bigg(\frac{r}{r_{\Sigma}}\bigg)\bigg\{4\sqrt{\varepsilon}\cos[\omega
(t-t_{0})]+b_{6}b_{7}(t)\ln\bigg(\frac{r}{r_{\Sigma}}\bigg)\bigg\}-3\varepsilon+3c_6^{2}c_7^{2}(t)+\\\nonumber&
2\sqrt{\varepsilon}\ln\bigg(\frac{r}{r_{\Sigma}}\bigg)\bigg\{\frac{\ddot{b_{7}}(t)}{b_{6}}\cos[\omega(t-t_{0})]+
3\dot{b_{7}}(t)\sin[\omega(t-t_{0})]\bigg\}+\bigg\{\ln\bigg(\frac{r}{r_{\Sigma}}\bigg)\bigg\}^{2}
\bigg\{2b_{7}(t)\ddot{b_{7}}(t)-3\dot{c_7}^{2}(t)\bigg\}\bigg]\\\nonumber&-\frac{1}{8\pi}\bigg[\frac{f-Rf_{R}}{2}+
\bigg\{\frac{\sqrt{\varepsilon}\cos[\omega(t-t_{0})]}{b_{6}}+b_{7}(t)\ln\bigg(\frac{r}{r_{\Sigma}}\bigg)\bigg\}^{2}
\bigg\{\ddot{f_{R}}-b_{6}\frac{-\sqrt{\varepsilon}\sin[\omega(t-t_{0})]+\dot{b_{7}}(t)\ln\bigg(\frac{r}
{r_{\Sigma}}\bigg)}{\sqrt{\varepsilon}\cos[\omega(t-t_{0})]+b_{6}b_{7}(t)\ln\bigg(\frac{r}{r_{\Sigma}}\bigg)}
\\\label{149}&\dot{f_{R}}\bigg\}
+3rb_{6}b_{7}(t)\bigg\{\sqrt{\varepsilon}\cos[\omega(t-t_{0})]+
b_{6}b_{7}(t)\ln\bigg(\frac{r}{r_{\Sigma}}\bigg)\bigg\}f'_{R}\bigg],
\end{align}
\begin{align}\nonumber
P_{\bot}=&\frac{f_{R}}{8\pi}\bigg[\varepsilon\cos^{2}[\omega(t-t_{0})]-2\sqrt{\varepsilon}b_{6}b_{7}(t)
\ln\bigg(\frac{r}{r_{\Sigma}}\bigg)\cos[\omega(t-t_{0})]-3\varepsilon+3b_{6}^{2}b_{7}^{2}(t)-\frac{2\sqrt{\varepsilon}}
{b_{6}}\ddot{b}_7(t)\ln\bigg(\frac{r}{r_{\Sigma}}\bigg)\\\nonumber&\cos[\omega(t-t_{0})]
+\bigg(\ln\bigg(\frac{r}{r_{\Sigma}}\bigg)
\bigg)^{2}\bigg\{2b_{7}(t)\ddot{b}_7(t)-3{\dot{b}_7}^{2}(t)\bigg\}+6\sqrt{\varepsilon}\dot{b}_7(t)\ln\bigg(\frac{r}
{r_{\Sigma}}\bigg)\sin[\omega(t-t_{0})]\bigg]-\frac{1}{8\pi}\bigg[\frac{f-Rf_{R}}{2}\\\nonumber&
+\bigg\{\frac{\sqrt{\varepsilon}\cos[\omega(t-t_{0})]}{b_{6}}+b_{7}(t)\ln\bigg(\frac{r}{r_{\Sigma}}\bigg)\bigg\}^{2}
\bigg\{\ddot{f_{R}}-b_{6}^{2}r^{2}f''_{R}\bigg\}-\bigg\{\frac{\sqrt{\varepsilon}\cos[\omega(t-t_{0})]}{b_{6}}+b_{7}(t)
\ln\bigg(\frac{r}{r_{\Sigma}}\bigg)\bigg\}\\\label{150}&\bigg\{-\sqrt{\varepsilon}\sin[\omega(t-t_{0})]+\dot{b}_7(t)\ln
\bigg(\frac{r}{r_{\Sigma}}\bigg)\bigg\}\dot{f_{R}}-b_{6} r\bigg\{\sqrt{\varepsilon}\cos[\omega(t-t_{0})]
+b_{6}b_{7}(t)\ln\bigg(\frac{r}{r_{\Sigma}}\bigg)\bigg\}(1+b_{7}(t))f'_{R}\bigg].
\end{align}
The physical variables corresponding to $C^{(10)}_\Sigma$ are given as
\begin{align}\nonumber
\mu=&\frac{f_{R}}{8\pi}\bigg[-2\varepsilon\sin^{2}[\omega(t-t_{0})]+b_{6}b_{7}(t)\ln\bigg(\frac{r}{r_{\Sigma}}\bigg)
\bigg\{b_{6}b_{7}(t)\ln\bigg(\frac{r}{r_{\Sigma}}\bigg)+2\sqrt{\varepsilon}\sin[\omega(t-t_{0})]\bigg\}\\\nonumber&
+3\varepsilon-3b_{6}^{2}b_{7}^{2}(t)+3{\dot{c_7}^{2}(t)}\bigg\{\ln\bigg(\frac{r}{r_{\Sigma}}\bigg)\bigg\}^{2}
+6\sqrt{\varepsilon}\dot{b_{7}}(t)\cos[\omega(t-t_{0})]\ln\bigg(\frac{r}{r_{\Sigma}}\bigg)\bigg]-\frac{1}{8\pi}
\\\nonumber&\bigg[\bigg\{r^{2}\bigg(\sqrt{\varepsilon}\sin[\omega(t-t_{0})]+b_{6}b_{7}(t)\ln\bigg(\frac{r}
{r_{\Sigma}}\bigg)\bigg)^{2}\bigg\}\bigg\{f''_{R}+\bigg(\frac{1}{r}-\frac{b_{6}b_{7}(t)}{r\bigg\{\sqrt{\varepsilon}
\sin[\omega(t-t_{0})]+b_{6}b_{7}(t)\ln\bigg(\frac{r}{r_{\Sigma}}\bigg)\bigg\}}\bigg)\\\label{151}&f'_{R}\bigg\}
\frac{3}{b_{6}}\bigg\{\sqrt{\varepsilon}\cos[\omega(t-t_{0})]+\dot{b_{7}}(t)\ln\bigg(\frac{r}{r_{\Sigma}}\bigg)\bigg\}
\bigg\{\sqrt{\varepsilon}\sin[\omega(t-t_{0})]+b_{6}b_{7}(t)\ln\bigg(\frac{r}{r_{\Sigma}}\bigg)\bigg\}\dot{f_{R}}-
\frac{f-Rf_{R}}{2}\bigg],
\end{align}
\begin{align}\nonumber
P_{r}=&\frac{f_{R}}{8\pi}\bigg[-b_{6}b_{7}(t)\ln\bigg(\frac{r}{r_{\Sigma}}\bigg)\bigg\{4\sqrt{\varepsilon}\sin[\omega
(t-t_{0})]+b_{6}b_{7}(t)\ln\bigg(\frac{r}{r_{\Sigma}}\bigg)\bigg\}-3\varepsilon+3c_6^{2}c_7^{2}(t)+\\\nonumber&
2\sqrt{\varepsilon}\ln\bigg(\frac{r}{r_{\Sigma}}\bigg)\bigg\{\frac{\ddot{b_{7}}(t)}{b_{6}}\sin[\omega(t-t_{0})]-
3\dot{b_{7}}(t)\cos[\omega(t-t_{0})]\bigg\}+\bigg\{\ln\bigg(\frac{r}{r_{\Sigma}}\bigg)\bigg\}^{2}
\bigg\{2b_{7}(t)\ddot{b_{7}}(t)-3\dot{c_7}^{2}(t)\bigg\}\bigg]\\\nonumber&-\frac{1}{8\pi}\bigg[\frac{f-Rf_{R}}{2}+
\bigg\{\frac{\sqrt{\varepsilon}\sin[\omega(t-t_{0})]}{b_{6}}+b_{7}(t)\ln\bigg(\frac{r}{r_{\Sigma}}\bigg)\bigg\}^{2}
\bigg\{\ddot{f_{R}}-b_{6}\frac{\sqrt{\varepsilon}\cos[\omega(t-t_{0})]+\dot{b_{7}}(t)\ln\bigg(\frac{r}
{r_{\Sigma}}\bigg)}{\sqrt{\varepsilon}\sin[\omega(t-t_{0})]+b_{6}b_{7}(t)\ln\bigg(\frac{r}{r_{\Sigma}}\bigg)}
\\\label{152}&\dot{f_{R}}\bigg\}
+3rb_{6}b_{7}(t)\bigg\{\sqrt{\varepsilon}\sin[\omega(t-t_{0})]+
b_{6}b_{7}(t)\ln\bigg(\frac{r}{r_{\Sigma}}\bigg)\bigg\}f'_{R}\bigg],
\end{align}
\begin{align}\nonumber
P_{\bot}=&\frac{f_{R}}{8\pi}\bigg[\varepsilon\sin^{2}[\omega(t-t_{0})]-2\sqrt{\varepsilon}b_{6}b_{7}(t)
\ln\bigg(\frac{r}{r_{\Sigma}}\bigg)\sin[\omega(t-t_{0})]-3\varepsilon+3b_{6}^{2}b_{7}^{2}(t)+\frac{2\sqrt{\varepsilon}}
{b_{6}}\ddot{b}_7(t)\ln\bigg(\frac{r}{r_{\Sigma}}\bigg)\\\nonumber&\sin[\omega(t-t_{0})]
+\bigg(\ln\bigg(\frac{r}{r_{\Sigma}}\bigg)
\bigg)^{2}\bigg\{2b_{7}(t)\ddot{b}_7(t)-3{\dot{b}_7}^{2}(t)\bigg\}-6\sqrt{\varepsilon}\dot{b}_7(t)\ln\bigg(\frac{r}
{r_{\Sigma}}\bigg)\cos[\omega(t-t_{0})]\bigg]-\frac{1}{8\pi}\bigg[\frac{f-Rf_{R}}{2}\\\nonumber&
+\bigg\{\frac{\sqrt{\varepsilon}\sin[\omega(t-t_{0})]}{b_{6}}+b_{7}(t)\ln\bigg(\frac{r}{r_{\Sigma}}\bigg)\bigg\}^{2}
\bigg\{\ddot{f_{R}}-b_{6}^{2}r^{2}f''_{R}\bigg\}-\bigg\{\frac{\sqrt{\varepsilon}\sin[\omega(t-t_{0})]}{b_{6}}+b_{7}(t)
\ln\bigg(\frac{r}{r_{\Sigma}}\bigg)\bigg\}\\\label{153}&\bigg\{\sqrt{\varepsilon}\cos[\omega(t-t_{0})]+\dot{b}_7(t)\ln
\bigg(\frac{r}{r_{\Sigma}}\bigg)\bigg\}\dot{f_{R}}-b_{6} r\bigg\{\sqrt{\varepsilon}\sin[\omega(t-t_{0})]
+b_{6}b_{7}(t)\ln\bigg(\frac{r}{r_{\Sigma}}\bigg)\bigg\}(1+b_{7}(t))f'_{R}\bigg].
\end{align}
Here, we have used the relation $\omega^{2}=c^{2}_{6}$. These solutions correspond to the situation where the CKV is parallel to the four-velocity and the heat flux $q=0$. The function $k(t)$ is formulated when the matching conditions \eqref{31} and \eqref{32} are satisfied. The differential Eq. \eqref{131}, which is obtained by the fulfillment of junction conditions, is integrated by introducing the different values of the parameters into it. The function $h(r)$ is assumed and the other functions are specified by imposing the diminishing complexity factor condition \eqref{29}. In this subsection, we have constructed four models. These models show the additional dark source terms as the last terms on the right side of the Equations, surrounded in square brackets.

\subsection{Case of $\chi^{\upsilon}\|V^{\upsilon};q\neq0$}

Ultimately, we explore the system which is dissipative and for which 4-velocity is $\|$ to CKV. For this case, the metric coefficients take the form as given in Eqs. \eqref{93}-\eqref{95}. Utilizing these results in Eq. A(3), we find
\begin{align}\label{154}
\frac{\dot{A'}}{A^{2}}-2\frac{\dot{A}A'}{A^{3}}=\frac{4\pi}{f_{R}}\bigg[qB-
\frac{1}{8\pi A}\bigg(\dot{f'_{R}}-\frac{A'}{A}\dot{f_{R}}-\frac{\dot{B}}{B}f'_{R}\bigg)\bigg].
\end{align}
The integration of the Eq. \eqref{154} produces
\begin{align}\label{155}
A=\frac{1}{g(r)+k(t)-4\pi\int\int I_2(r,t)drdt},
\end{align}
\begin{align}\label{156}
B=\frac{1}{l(r)[g(r)+k(t)-4\pi\int\int I_2(r,t)drdt]},
\end{align}
\begin{align}\label{157}
C=\frac{r}{l(r)[g(r)+k(t)-4\pi\int\int I_2(r,t)drdt]},
\end{align}
where $k(t)$ and $g(r)$ are integration entities. Here, the expression of $I_2(r,t)$ is mentioned in Appendix $A$. To find these arbitrary functions, we apply vanishing complexity factor condition $(Y_{TF}=0)$.
By utilizing Eq.s \eqref{155}-\eqref{157}, Eq. \eqref{154} takes the form
\begin{align}\label{158}
g'-4\pi\int I_4(r,t)dt=\frac{\lambda(t)r}{l^{2}(r)},
\end{align}
which produces
\begin{align}\label{159}
g-4\pi\int\int I_4(r,t)dtdr=\lambda(t)\int\frac{rdr}{l^{2}(r)},
\end{align}
with $\lambda(t)$ as an arbitrary function. The differentiation of Eq. \eqref{158} with regard to time gives
\begin{align}\label{160}
4\pi I_4(r,t)=\frac{-\dot{\lambda}(t)r}{l^{2}(r)}.
\end{align}
Using this result, the metric coefficients may be formulated as
\begin{align}\label{161}
B=\frac{1}{l(r)\bigg[k(t)+\lambda(t)\int\frac{rdr}{l^2(r)}\bigg]},\quad\quad A=\frac{1}{\bigg[k(t)+\lambda(t)\int\frac{rdr}{l^2(r)}\bigg]},\quad\quad C=\frac{r}{l(r)\bigg[k(t)+\lambda(t)\int\frac{rdr}{l^2(r)}\bigg]}.
\end{align}
In order to specify other functions $k(t)$, $\lambda(t)$ we will impose junction condition $(q=P_{r}+\frac{f-Rf_{R}}{16\pi})_{\Sigma}$, which produces
\begin{align}\label{162}
l^{2}_{\Sigma}H_{\Sigma}\bigg(\frac{1}{r_{\Sigma}}-\frac{l'_\Sigma}{l_\Sigma}-\frac{H'_\Sigma}{H_\Sigma}\bigg)
\bigg(\frac{1}{r_{\Sigma}}-\frac{l'_\Sigma}{l_\Sigma}-3\frac{H'_\Sigma}{H_\Sigma}\bigg)-H_\Sigma\bigg(
-2\frac{\ddot{H_\Sigma}}{H_\Sigma}+3\frac{\dot{H^{2}_\Sigma}}{H^{2}_\Sigma}+\eta^{2}\bigg)=
-\frac{2\dot{\lambda}}{\eta},
\end{align}
where
\begin{align}\label{163}
H\equiv k(t)+\lambda(t)\int\frac{rdr}{l^{2}(r)},\quad\quad \eta\equiv\frac{l_\Sigma}{r_{\Sigma}}.
\end{align}
To find the solution of Eq. \eqref{162}, we suppose
\begin{align}\label{164}
-2\frac{\ddot{H_\Sigma}}{H_\Sigma}+3\frac{\dot{H^{2}_\Sigma}}{H^{2}_\Sigma}+\eta^{2}=
-\frac{2\dot{\lambda}}{H_\Sigma\eta},\quad\quad 1-b_{1}-\frac{r_{\Sigma} H'_\Sigma}{H_\Sigma}=0,
\end{align}
which gives
\begin{align}\label{165}
\frac{\lambda(t)\dot{H_\Sigma}}{H^{2}_\Sigma}=\frac{\dot{\lambda}(t)}{H_\Sigma}.
\end{align}
Equation \eqref{164} can be written as
\begin{align}\label{166}
2\frac{\ddot{H_\Sigma}}{H_\Sigma}-3\frac{\dot{H^{2}_\Sigma}}{H^{2}_\Sigma}+\frac{2\eta\dot{H_\Sigma}(1-b_{1})}
{H_\Sigma}-\eta^{2}=0.
\end{align}
To find the solution of Eq. \eqref{166}, we propose $v=\frac{\dot{H}_\Sigma}{H_\Sigma}$. Thus, the Eq. \eqref{166} reads
\begin{align}\label{167}
\dot{v}-\frac{1}{2}v^{2}+\eta(1-b_{1})v-\frac{\eta^{2}}{2}=0.
\end{align}
This equation is known as the Ricatti equation. A particular solution of this Eq. is
\begin{align}\label{168}
v_0=\eta(1-b_{1})\pm\eta\sqrt{b_{1}^{2}-2b_{1}}.
\end{align}
And the general solution of this Ricatti equation can be found by substituting $y=v-v_0$ in Eq. \eqref{167}, giving us
\begin{align}\label{169}
\dot{y}-\frac{1}{2}y^{2}+[\eta(1-b_{1})-v_0]y=0.
\end{align}
The solution of Eq. \eqref{169} is
\begin{align}\label{170}
y=\frac{2\nu}{1+be^{\nu t}},
\end{align}
here $b$ is integration constant and $\nu\equiv\mu(1-b_{1})-v_0$. The final expression $H_\Sigma$ is
\begin{align}\label{171}
H_\Sigma=\frac{ae^{(2\nu+v_0)t}}{(1+be^{\nu t})^{2}},
\end{align}
where $a$ is an integration constant. The physical variables in the form of functions $h(r)$, and $H(t,r)$ read as
\begin{align}\nonumber
8\pi\mu=&f_{R}\bigg[3\dot{H^{2}}-l^{2}H^{2}\bigg\{\frac{-2l''}{l}+\frac{3l'^{2}}{l^{2}}-2\frac{H''}{H}+
3\frac{H'^{2}}{H^{2}}+2\frac{l'}{l}\frac{H'}{H}-\frac{4}{r}\frac{l'}{l}-\frac{4}{r}\frac{H'}{H}\bigg\}\bigg]
\\\label{172}&-\bigg\{f''_{R}l^{2}H^{2}+l^{2}H^{2}\bigg(\frac{l'}{l}-\frac{H'}{H}+\frac{2}{r}\bigg)f'_{R}+3H\dot{H}\dot{f_{R}}-
\frac{f-Rf_{R}}{2}\bigg\},\\\label{173}
8\pi q&=-2lH\dot{H'}f_{R}+lH^{2}\bigg[\dot{f'_{R}}-\bigg(\frac{2l'}{l}-\frac{H'}{H}\bigg)
\dot{f_{R}}+\frac{\dot{H}}{H}f'_{R}\bigg],\\\nonumber
8\pi P_{r}=&f_{R}\bigg[2H\ddot{H}-3\dot{H^{2}}+l^{2}H^{2}\bigg\{\frac{l'^{2}}{l^{2}}+\frac{3H'^{2}}{H^{2}}+
\frac{4l'}{l}\frac{H'}{H}-\frac{2l'}{rl}-\frac{4H'}{rH}\bigg\}\bigg]\\\label{174}&
-\bigg\{\frac{f-Rf_{R}}{2}+H^{2}\ddot{f_{R}}-H\dot{H}\dot{f_{R}}-l^{2}H^{2}\bigg(\frac{4l'}{l}-\frac{3H'}{H}+
\frac{2}{r}\bigg)f'_{R}\bigg\},\\\nonumber
8\pi P_{\bot}=&f_{R}\bigg[2H\ddot{H}-3\dot{H^{2}}+l^{2}H^{2}\bigg\{\frac{-l''}{l}+\frac{l'^{2}}{l^{2}}-
\frac{2H''}{H}+\frac{3H'^{2}}{H^{2}}-\frac{l'}{rl}-\frac{2H'}{rH}\bigg\}\bigg]\\\label{175}&
-\bigg\{\frac{f-Rf_{R}}{2}+H^{2}\ddot{f_{R}}-l^{2}H^{2}f''_{R}-2H\dot{H}\dot{f_{R}}+
l^{2}H^{2}\bigg(\frac{2l'}{l}-\frac{H'}{H}-
\frac{1}{r}\bigg)f'_{R}\bigg\}.
\end{align}
To find a particular model, we suppose that $b_{1}=2$, which gives $v_0=-\eta$ and $\nu=0$, thus $X_{\Sigma}$ read $X_{\Sigma}=c e^{-\eta t}$,where $c=\frac{a}{(1+b)^{2}}$. Further, we assume that $h(r)=c_{2}r^{2}$, with $c_{2}$ is a constant having dimension $\frac{1}{[length]^{2}}$, giving us
$\eta=c_{2}r_{\Sigma}$. Using these relations, we obtain
\begin{align}\label{176}
\int\frac{rdr}{l^{2}(r)}=-\frac{r^{2}_{\Sigma}}{2\eta^{2}r^{2}},\quad\quad \lambda(t)=-\eta^{2}ce^{-\eta t}.
\end{align}
Thus, the function $k(t)$ is formulated as $k(t)=\frac{ce^{-\eta t}}{2}$. The final expression for $H(t,r)$ is
\begin{align}\label{177}
H^{(11)}(t,r)=\frac{ce^{-\eta t}}{2}\bigg[1+\bigg(\frac{r_{\Sigma}}{r}\bigg)^{2}\bigg].
\end{align}
The physical variables in the form of $H^{(11)}(t,r)$ read
\begin{align}\nonumber
&8\pi\mu=\frac{3c^{2}\eta^{2}e^{-2\eta t}}{4r^{4}}\bigg(5r^4+2r^{2}r_{\Sigma}^{2}+r_{\Sigma}^{4}\bigg)f_{R}-
\bigg[\frac{c_{2}c^{2}e^{-2\eta t}}{4}(r^{2}+ r_{\Sigma}^{2})+\bigg\{f''_{R}+\bigg(\frac{4}{r}+
\frac{2r_{\Sigma}^{2}}{r(r^{2}+r_{\Sigma}^{2})}\bigg)f'_{R}\bigg\}\\\label{178}&-\frac{3}{4r^{2}}\eta c^{2}e^{-2\eta t}(r^{2}+r_{\Sigma}^{2})\dot{f_{R}}-\frac{f-Rf_{R}}{2}\bigg],\\\label{179}
&8\pi q=-c^{2}\eta^{2}e^{-2\eta t}(r^{2}+ r_{\Sigma}^{2})\frac{r_{\Sigma}}{r^{3}}f_{R}+\frac{c_{2}c^{2}e^{-2\eta t}}{4r^2}(r^{2}+r_{\Sigma}^{2})^{2}\bigg\{\dot{f'_{R}}-
\bigg(\frac{4}{r}+\frac{2r_{\Sigma}^{2}}{r(r^{2}+r_{\Sigma}^{2})}\bigg)\dot{f_{R}}-\eta f'_{R}\bigg\},\\\nonumber
&8\pi P_{r}=\frac{c^{2}\eta^{2}e^{-2\eta t}}{4r^{4}}(2r^{2} r_{\Sigma}^{2}-9r^{4}-r^{4}_\Sigma)f_{R}-
\bigg\{\frac{f-Rf_{R}}{2}+\frac{c^{2}e^{-2\eta t}}{4r^{4}}(r^{2}+r_{\Sigma}^{2})\ddot{f_{R}}\\\label{180}&
+\frac{\eta c^{2}e^{-2\eta t}}{4r^{4}}(r^{2}+r_{\Sigma}^{2})^{2}\dot{f_{R}}-\frac{c_{2}^{2} c^{2}e^{-2\eta t}}{4r^{2}}(r^{2}+r_{\Sigma}^{2})^{2}\bigg(\frac{16}{r}+
\frac{6r^{2}_\Sigma}{r(r^{2}+r_{\Sigma}^{2})}\bigg)f'_{R}\bigg\},\\\nonumber
&8\pi P_{\bot}=\frac{c^{2}\eta^{2}e^{-2\eta t}}{4r^{4}}(2r^{2} r_{\Sigma}^{2}-9r^{4}-r^{4}_{\Sigma})f_{R}-
\bigg\{\frac{f-Rf_{R}}{2}+\frac{c^{2}e^{-2\eta t}}{4r^{4}}(r^{2}+r_{\Sigma}^{2})\ddot{f_{R}}
-\frac{c_{2}^{2}c^{2}e^{-2\eta t}}{4}(r^{2}+r_{\Sigma}^{2})^{2}f''_{R}\\\label{181}&
+\frac{\eta c^{2}e^{-2\eta t}}{2r^{4}}(r^{2}+r_{\Sigma}^{2})^{2}\dot{f_{R}}+\frac{c_{2}^{2} c^{2}e^{-2\eta t}}{4}(r^{2}+r_{\Sigma}^{2})^{2}\bigg(\frac{3}{r}+\frac{2r^{2}_{\Sigma}}{r(r^{2}+
r_{\Sigma}^{2})}\bigg)f'_{R}\bigg\},
\end{align}
The corresponding total matter quantity ($m$) and the temperature turned out to be
\begin{align}\label{182}
&m_{\Sigma}=\frac{e^{\eta(t)}}{c\eta(t)},\\\label{183}
&\mathcal{T}(r,t)=\frac{ce^{-\eta(t)}}{2r^{2}}(r^{2}+r^{2}_\Sigma)\bigg[\int\bigg\{\frac{\tau cr_{\Sigma}\eta^{3}e^{-\eta(t)}}{2\pi c_{2}\mathcal{K} r^{3}}f_{R}+\frac{\eta^{2}}{2\pi c_{2}\mathcal{K} r(r^{2}+r^{2}_\Sigma)}f_{R}\bigg\}dr+\int Z_{2}(r,t)dr+{\mathcal{T}_{0}}(t)\bigg].
\end{align}
The metric coefficients for this case as mentioned in Eqs. \eqref{172}-\eqref{175}, which, after implying the diminishing complexity factor condition $(Y_{TF}=0)$, transforms as given in Eqs. \eqref{178}-\eqref{181}. Thus, the spacetime is determined up to four functions (two functions of $t$, one function of $r$, and one function of $t$ and $r$). The function of $t$ is acquired from the integration of the matching conditions \eqref{35} and \eqref{36}, whereas the function $g(r)$ is presumed along with the choice $a_{1}=2$. This produces the model \textbf{11}. The additional dark source terms resulting from the theory are shown as the last terms on the right side of the resulting field equations enclosed in square brackets in this model.

\section{conclusions}

The field equations for general spherically symmetric fluid configuration can be solved in many ways as a result of the acceptance of CKV, as we have seen earlier. The present study is recognized as a kind of generalization of Herrera's work \cite{herrera2022non}. We have found a number of solutions of field equations in $f(R)$ gravity. These solutions can be used to solve a wide range of astrophysical issues or as test theoretical models for the study of hypothetical concepts like white holes and wormholes. We added more restrictions to the fluid distribution in order to identify solutions that could be stated in regard of elementary functions. Some of these were imposed solely to create models represented by simple functions, while others (such as the diminishing complexity factor or the quasi-homologous condition) are implied with a clear physical meaning.

  In subsection \textbf{V A}, we found five models. The first two models are obtained by choosing $\alpha=\frac{1}{27M^{2}}$ giving us the values of areal radius $C^{(1)}_{\Sigma}$ and $C^{(2)}_{\Sigma}$, describing the expansion and contraction of the fluid from $0$ to $3M$ and from infinity to $3M$, respectively. These models show that the energy densities of the system are positive and the first model has only one singularity at $t=t_{0}$.  The third model is obtained by choosing $\alpha=0$, representing that the areal radius fluctuates between $0$ and $2M$. This model's radial pressure only depends on $f(R)$ curvature terms. In addition, we considered the case $M=0$, showing that the fluid distribution has no gravitational effect outside the boundary surface. These models represent a kind of ``ghost stars" but they demonstrate pathologies, both topological and physical. Hence their physical uses are unclear. These types of distributions have been studied in the past \cite{zeldovich1971relativistic}.

In subsection \textbf{V B} for the dissipative case, we have obtained the model \textbf{6}. The corresponding equations of motion modify in $f(R)$ gravity analysis. The symmetry assumed in subsection \textbf{V C}, again lowers the metric variables (two functions of $t$, one function of $r$, and one function of $t$ and $r$) to four functions. The model \textbf{7} is obtained by taking the particular value of constant $a_{1}$ ($a_{1}=1$) and taking $h(r)=c_{6}r$. The expression of $C^{(7)}_{\Sigma}$ shows that the areal radius expands from $0$ to $3M$. For this model, the valve of energy density is positive and is non-singular at $t=t_{0}$. The model \textbf{8} is obtained by taking the particular value of constants $a_{1}$ ($a_{1}=\frac{1}{2}$) and $\varepsilon$ ($\varepsilon=0$). The expression of $C^{(8)}_\Sigma$ shows that the areal radius oscillates between $0$ to $\frac{8M}{3}$. The value of $\mu$ is positive but the fluid configuration has a singularity at $r=0$. Models \textbf{9} and \textbf{10} are obtained for $M=0$ and $a_{1}=1$. They represent the type of ``ghost stars" that was previously discussed.

Finally, in subsection \textbf{V D}, we studied the dissipative case for the CKV parallel to the four-velocity. The model \textbf{11} is found by taking $a_{1}=2$. In this model, the value of the areal radius compensates for the decline in energy density and heat flux by rising exponentially. In general, all the obtained solutions (or some of them) may be helpful to mark out the evolution of some phases of the collapsing fluid in the framework of metric $f(R)$ gravity theory. The pertinent parameters are given particular values according to each particular case. It is important to remember that in any actual scenario involving a collapse, we cannot anticipate that the same equation of state would hold true throughout the evolution and for the entire fluid configuration. Secondly, the regularity conditions are not fulfilled at the center of the fluid configuration. Thirdly, if we take $f(R)=R$, all the obtained results are reduced to GR.

\section*{Acknowledgement}

The work of KB was supported in part by the JSPS KAKENHI Grant Number JP21K03547.

\section*{Data Availability Statement}

All data generated or analyzed during this study are included in this published article.

\section*{Appendix A: Field equations}

For the interior line element, the field equations in metric $f(R)$ gravity are as
\begin{align}\tag{A1}
G_{\rho\nu}=\frac{8\pi}{f_{R}}\bigg(\mathbb{T^{(\mathfrak{M})}_{\rho\nu}}+
\mathbb{T^{(\mathfrak{D})}_{\rho\nu}}\bigg),
\end{align}
and its nonzero components of field equations are
\begin{align}\nonumber
\mu=&\frac{f_{R}}{8\pi}\bigg[\frac{1}{A^{2}}\bigg(2\frac{\dot{B}}{B}+\frac{\dot{C}}{C}\bigg)\frac{\dot{C}}{C}
-\frac{1}{B^{2}}\bigg\{2\frac{C''}{C}+\bigg(\frac{C'}{C}\bigg)^{2}-2\frac{B'}{B}\frac{C'}{C}-\frac{1}{8\pi}  \bigg[\frac{f''_{R}}{B^{2}}
\bigg(\frac{B}{C}\bigg)^{2}\bigg\}\bigg]+\frac{1}{B^{2}}\bigg(2\frac{C'}{C}\\\tag{A2}&-\frac{B'}{B}\bigg)f'_{R}-\frac{1}{A^{2}}
\bigg(2\frac{\dot{C}}{C}+\frac{\dot{B}}{B}\bigg)\dot{f_{R}}-\frac{f-Rf_{R}}{2}\bigg],\\\tag{A3}\label{eq:A3}
q=&\frac{1}{AB}\bigg[\frac{f_{R}}{4\pi}\bigg(\frac{\dot{C'}}{C}-\frac{\dot{B}}{B}\frac{C'}{C}-
\frac{\dot{C}}{C}\frac{C'}{C}\bigg)+\frac{1}{8\pi}\bigg(\dot{f'_{R}}-\frac{A'}{A}\dot{f_{R}}-
\frac{\dot{B}}{B}f'_{R}\bigg)\bigg],
\\\nonumber
P_{r}=&\frac{f_{R}}{8\pi}\bigg[\frac{-1}{A^{2}}\bigg\{2\frac{\ddot{C}}{C}-\bigg(2\frac{\dot{A}}{A}-
\frac{\dot{C}}{C}\bigg)\frac{\dot{C}}{C}\bigg\}+\frac{1}{B^{2}}\bigg(2\frac{A'}{A}+
\frac{C'}{C}\bigg)\frac{C'}{C}-\frac{1}{C^{2}}\bigg]-\frac{1}{8\pi}\bigg[\frac{f-Rf_{R}}{2}
+\frac{1}{A^{2}}\ddot{f_{R}}\\\tag{A4}&+\frac{1}{A^{2}}\bigg(2\frac{\dot{C}}{C}-\frac{\dot{A}}{A}\bigg)\dot{f_{R}}-
\frac{1}{B^{2}}\bigg(\frac{A'}{A}+2\frac{C'}{C}\bigg)f'_{R}\bigg],
\\\nonumber
P_{\bot}=&\frac{f_{R}}{8\pi}\bigg[\frac{-1}{A^{2}}\bigg\{\frac{\ddot{B}}{B}+\frac{\ddot{C}}{C}-\frac{\dot{A}}{A}
\bigg(\frac{\dot{B}}{B}+\frac{\dot{C}}{C}\bigg)+\frac{\dot{B}}{B}\frac{\dot{C}}{C}\bigg\}+\frac{1}{B^{2}}
\bigg\{\frac{A''}{A}+\frac{C''}{C}-\frac{A'}{A}\frac{B'}{B}+\bigg(\frac{A'}{A}-\frac{B'}{B}\bigg)
\frac{C'}{C}\bigg\}\bigg]\\\tag{A5}&-\frac{1}{8\pi}\bigg[\frac{f-Rf_{R}}{2}+\frac{1}{A^{2}}\ddot{f_{R}}-
\frac{1}{B^{2}}f''_{R}+\frac{1}{A^{2}}\bigg(\frac{\dot{C}}{C}-\frac{\dot{A}}{A}+\frac{\dot{B}}{B}\bigg)
\dot{f_{R}}+\frac{1}{B^{2}}\bigg(\frac{A'}{A}+\frac{B'}{B}-\frac{C'}{C}\bigg)f'_{R}\bigg].
\end{align}
Using expansion and shear scalars, the component \eqref{eq:A3} is written as
\begin{align}\tag{A6}\label{eq:A6}
4\pi qB= f_{R}\bigg[\frac{1}{3}(\Theta-\sigma)'-\sigma\frac{C'}{C}\bigg]+\frac{1}{2A}\bigg[\dot{f'_{R}}-
\frac{A'}{A}\dot{f_{R}}-\frac{\dot{B}}{B}f'_{R}\bigg].
\end{align}
The arbitrary function appearing in the value of metric coefficients $A$ and $B$ of Eq. \eqref{52} is found as under
\begin{align}\tag{A7}
I(r,t)=-\int\int\frac{1}{2Bf_{R}}\bigg[\dot{f'_{R}}-\frac{B'}{B}\dot{f_{R}}-\frac{\dot{B}}{B}f'_{R}\bigg]drdt.
\end{align}
The value of the temperature function $\mathcal{T}(r,t)$ under the conditions $\chi_{\upsilon}V^{\upsilon}=0$ and $q\neq0$ as shown in Eq. \eqref{122}. Its value is found as under
\begin{align}\nonumber
Z_{1}(r,t)=&\frac{\tau\omega}{4\pi\mathcal{K}\delta(t)(\omega r+1)}\bigg[\frac{2\delta(t)\dot{\delta}(t)(\omega r+1)^{2}}{8\pi\omega^{2}c_{3}}\bigg\{\dot{f'_{R}}+\frac{\omega}{\omega r+1}\dot{f_{R}}+\frac{\dot{\delta}(t)}{{\delta}(t)}f'_{R}\bigg\}+\frac{{\delta}^{2}(t)(\omega r+1)^{2}}{8\pi\omega^{2}c_{3}}\bigg\{\ddot{f'_{R}}+\frac{\omega}{\omega r+1}\ddot{f_{R}}+\frac{\ddot{\delta}(t)}{{\delta}(t)}\\\tag{A8}&\times f'_{R}-\frac{\dot{\delta}^{2}(t)}{{\delta}^{2}(t)}f'_{R}
+\frac{\dot{\delta}(t)}{{\delta}(t)}\dot{f'}_R\bigg\}\bigg]-\bigg[\frac{c_{3}{\omega}^{2}}{4\pi
\mathcal{K}\delta(t)(\omega r+1)^{2}}\bigg\{
\frac{{\delta}^{2}(t)(\omega r+1)^{2}}{8\pi c_{3}{\omega}^{2}}\bigg(\dot{f'_{R}}+\frac{\omega}{(\omega r+1)}\dot{f_{R}}+\frac{\dot{\delta}}{\delta}f'_{R}\bigg)\bigg\}\bigg].
\end{align}
The metric coefficients $A$ under the conditions $\chi_{\upsilon}V^{\upsilon}=0$ and $q\neq0$ contains $I_2$ as an arbitrary function as shown in Eq. \eqref{97}. Its value is found as under
\begin{align}\tag{A9}
I_{2}(r,t)=-\int\int\frac{1}{2Af_{R}}\bigg[\dot{f'_{R}}-\frac{A'}{A}\dot{f_{R}}-\frac{\dot{A}}{A}f'_{R}\bigg]drdt.
\end{align}
The temperature function within the background of $\chi^{\upsilon}\|V^{\upsilon}$ and $q\neq0$ contains $Z_{2}(r,t)$ as shown in Eq. \eqref{183}. Its value is found as under
\begin{align}\nonumber
Z_{2}(r,t)=&\frac{\tau\eta ce^{-\eta(t)}(r^{2}+r^{2}_\Sigma)}{8\pi\mathcal{K} r^{2}}\bigg[\dot{f'_{R}}-\bigg\{
\frac{4}{r}+\frac{2r^{2}_\Sigma}{r({r^{2}+r^{2}_\Sigma})}\bigg\}\dot{f_{R}}-\eta f'_{R}\bigg]
-\frac{c_{2}c^{2}e^{-2\eta(t)}}{32\pi r^{2}}({r^{2}+r^{2}_\Sigma})\bigg[\ddot{f'_{R}}-\bigg\{
\frac{4}{r}+\frac{2r^{2}_\Sigma}{r({r^{2}+r^{2}_\Sigma})}\bigg\}\\\tag{A10}&\times \ddot{f_{R}}-\eta \dot{f'_{R}}\bigg]\frac{1}{8\pi \mathcal{K}}\bigg[\dot{f'_{R}}-\bigg\{
\frac{4}{r}+\frac{2r^{2}_\Sigma}{r({r^{2}+r^{2}_\Sigma})}\bigg\}\dot{f_{R}}-\eta f'_{R}\bigg].
\end{align}

\vspace{0.5cm}

\section*{Appendix B}

The physical variables corresponding to $C^{(1)}_\Sigma$ described in Eq. \eqref{69} are found as under
\begin{align}\nonumber
\mu&=\frac{f_{R}}{8\pi}\bigg[-3c_{1}^{2}(t)+\frac{\omega^{2}}{F_{0}^{2}}\bigg\{\frac{F_{0}}{6\omega M}\bigg(3\coth^{2}[\frac{\omega}{2}(t-t_{0})]-1\bigg)-c_{1}(r_{\Sigma}-r)\bigg\}^{2}+\frac{3}{F_{0}^{2}}
\bigg\{\frac{-F_{0}}{2M}\frac{\cosh[\frac{\omega}{2}(t-t_{0})]}{\sinh^{3}[\frac{\omega}{2}
(t-t_{0})]}\\\nonumber
&-\dot{c_{1}}(t)(r_{\Sigma}-r)\bigg\}^{2}\bigg]-\frac{1}{8\pi}\bigg[\bigg\{\frac{F_{0}}
{6\omega M}\bigg(3\coth^{2}[\frac{\omega}{2}(t-t_{0})]-1\bigg)-c_{1}(t)(r_{\Sigma}-r)\bigg\}^{2}f''_{R}
\\\nonumber
&-\bigg\{\frac{F_{0}}{6\omega M}\bigg(3\coth^{2}[\frac{\omega}{2}(t-t_{0})]-1\bigg)-c_{1}(t)(r_{\Sigma}-r)\bigg\}c_{1}(t)f'_{R}
+\frac{1}{F_{0}}\bigg\{\frac{F_{0}}{6\omega M}\bigg(3\coth^{2}[\frac{\omega}{2}(t-t_{0})]-1\bigg)\\\label{70}
&-c_{1}(r_{\Sigma}-r)\bigg\}
\bigg\{\frac{-F_{0}}{2M}\frac{\cosh[\frac{\omega}{2}(t-t_{0})]}{\sinh^{3}[\frac{\omega}{2}
(t-t_{0})]}-\dot{c_{1}}(t)(r_{\Sigma}-r)\bigg\}\dot{f_{R}}-\frac{f-Rf_{R}}{2}\bigg],\\\nonumber
P_{r}&=\frac{f_{R}}{8\pi}\bigg[\frac{2\omega c_{1}(t)(r_{\Sigma}-r)}{3F_{0}M}-\frac{\omega^{2}c_{1}^{2}(t)
(r_{\Sigma}-r)^{2}}{F_{0}^{2}}-\frac{3\omega c_{1}(t)(r_{\Sigma}-r)}{2F_{0}M\sinh^{4}[\frac{\omega}{2}(t-t_{0})]}-
\frac{3}{F_{0}^{2}}\bigg\{\dot{c_{1}}^{2}(t)(r_{\Sigma}-r)^{2}+\frac{F_{0}\dot{c_{1}}(t)
(r_{\Sigma}-r)}{M}\\\nonumber
&\frac{\cosh[\frac{\omega}{2}(t-t_{0})]}{\sinh^{3}[\frac{\omega}{2}(t-t_{0})]}\bigg\}-\frac{2\ddot{c_{1}}(t)
(r_{\Sigma}-r)}{F_{0}^{2}}\bigg\{\frac{F_{0}}{6\omega M}\bigg(3\coth^{2}[\frac{\omega}{2}(t-t_{0})]-1\bigg)
-c_{1}(t)(r_{\Sigma}-r)\bigg\}\bigg]-\frac{1}{8\pi}\bigg[\frac{f-Rf_{R}}
{2}-\\\nonumber&\frac{1}{F_{0}^{2}}\bigg\{\frac{F_{0}}{6\omega M}\bigg(3\coth^{2}[\frac{\omega}{2}(t-t_{0})]-1\bigg)
-c_{1}(t)(r_{\Sigma}-r)\bigg\}^{2}\ddot{f_{R}}-\frac{1}{F_{0}^{2}}\bigg\{\frac{F_{0}}{6\omega M}\bigg(3\coth^{2}[\frac{\omega}{2}(t-t_{0})]-1\bigg)-\\\nonumber&c_{1}(t)(r_{\Sigma}-r)\bigg\}
\bigg\{\frac{-F_{0}}{2M}\frac{\cosh[\frac{\omega}{2}(t-t_{0})]}{\sinh^{3}[\frac{\omega}{2}
(t-t_{0})]}-\dot{c_{1}}(t)(r_{\Sigma}-r)\bigg\}\dot{f_{R}}+3c_{1}(t)\bigg\{\frac{F_{0}}{6\omega M}\bigg(3\coth^{2}[\frac{\omega}{2}(t-t_{0})]-1\bigg)-\\\label{71}&c_{1}(t)(r_{\Sigma}-r)\bigg\}f'_{R}\bigg],\\\nonumber
P_{\bot}&=\frac{f_{R}}{8\pi}\bigg[3c_{1}^{2}(t)+\frac{[F_{0}-3\omega c_{1}M(r_{\Sigma}-r)]}{3F_{0}M^{2}\sinh^{2}[\frac{\omega}{2}(t-t_{0})]}+\frac{[F_{0}-6\omega c_{1}M(r_{\Sigma}-r)]}{4F_{0}M^{2}\sinh^{4}[\frac{\omega}{2}(t-t_{0})]}-\frac{3}{F_{0}^{2}}\bigg\{\dot{c_{1}^{2}}(t)
(r_{\Sigma}-r)^{2}+\frac{F_{0}\dot{c_{1}}(t)(r_{\Sigma}-r)}{M}\\\nonumber&\frac{\cosh[\frac{\omega}{2}(t-t_{0})]}{\sinh^{3}
[\frac{\omega}{2}(t-t_{0})]}\bigg\}-\frac{3}{F_{0}^{2}}\bigg\{\frac{F_{0}^{2}}{4M^{2}}\frac{\cosh^{2}[\frac{\omega}
{2}(t-t_{0})]}{\sinh^{6}[\frac{\omega}{2}(t-t_{0})]}+\dot{c_{1}^{2}}(t)(r_{\Sigma}-r)^{2}+\frac{F_{0}\dot{c_{1}}(t)
(r_{\Sigma}-r)}{M}\frac{\cosh[\frac{\omega}{2}(t-t_{0})]}{\sinh^{3}[\frac{\omega}{2}(t-t_{0})]}\bigg\}\bigg]
\\\nonumber&-\frac{1}{8\pi}\bigg[\frac{f-Rf_{R}}{2}+\bigg\{\frac{F_{0}}{6\omega M}\bigg(3\coth^{2}[\frac{\omega}{2}(t-t_{0})]-1\bigg)-c_{1}(t)(r_{\Sigma}-r)\bigg\}\bigg\{\frac{1}{F_{0}^{2}}
\ddot{f_{R}}+f''_{R}\bigg\}\\\nonumber&+\bigg\{\frac{F_{0}}{6\omega M}\bigg(3\coth^{2}[\frac{\omega}{2}(t-t_{0})]-1\bigg)
-c_{1}(t)(r_{\Sigma}-r)\bigg\}\bigg\{\frac{1}{F_{0}^{2}}\bigg(\frac{-F_{0}}{2M}\frac{\cosh[\frac{\omega}{2}(t-t_{0})]}{\sinh^{3}[\frac{\omega}{2}
(t-t_{0})]}-\dot{c_{1}}(t)(r_{\Sigma}-r)\bigg)\dot{f_{R}}-\\\label{72}&c_{1}(t)f'_{R}\bigg\}\bigg].
\end{align}
The above-mentioned results are obtained in the background of $\chi_{\upsilon}V^{\upsilon}=0$ constraint for the adiabatic spherical structure.

The physical variables corresponding to $C^{(2)}_\Sigma$ described in Eq.\eqref{74} are
\begin{align}\nonumber
\mu&=\frac{f_{R}}{8\pi}\bigg[-3c_{1}^{2}(t)+\frac{\omega^{2}}{F_{0}^{2}}\bigg\{\frac{F_{0}}{6\omega M}\bigg(3\tanh^{2}[\frac{\omega}{2}(t-t_{0})]-1\bigg)-c_{1}(r_{\Sigma}-r)\bigg\}^{2}+\frac{3}{F_{0}^{2}}
\bigg\{\frac{F_{0}}{2M}\frac{\sinh[\frac{\omega}{2}(t-t_{0})]}{\cosh^{3}[\frac{\omega}{2}
(t-t_{0})]}-\\\nonumber&\dot{c_{1}}(t)(r_{\Sigma}-r)\bigg\}^{2}\bigg]-\frac{1}{8\pi}\bigg[\bigg\{\frac{F_{0}}
{6\omega M}\bigg(3\tanh^{2}[\frac{\omega}{2}(t-t_{0})]-1\bigg)-c_{1}(t)(r_{\Sigma}-r)\bigg\}^{2}f''_{R}-\\\nonumber&
\bigg\{\frac{F_{0}}{6\omega M}\bigg(3\tanh^{2}[\frac{\omega}{2}(t-t_{0})]-1\bigg)-c_{1}(t)(r_{\Sigma}-r)\bigg\}c_{1}(t)f'_{R}
+\frac{1}{F_{0}}
\bigg\{\frac{F_{0}}{6\omega M}\bigg(3\tanh^{2}[\frac{\omega}{2}(t-t_{0})]-1\bigg)-\\\label{75}&c_{1}(r_{\Sigma}-r)\bigg\}
\bigg\{\frac{-F_{0}}{2M}\frac{\sinh[\frac{\omega}{2}(t-t_{0})]}{\cosh^{3}[\frac{\omega}{2}
(t-t_{0})]}-\dot{c_{1}}(t)(r_{\Sigma}-r)\bigg\}\dot{f_{R}}-\frac{f-Rf_{R}}{2}\bigg],\\\nonumber
P_{r}&=\frac{f_{R}}{8\pi}\bigg[\frac{2\omega c_{1}(t)(r_{\Sigma}-r)}{3F_{0}M}-\frac{\omega^{2}c_{1}^{2}(t)
(r_{\Sigma}-r)^{2}}{F_{0}^{2}}-\frac{3\omega c_{1}(t)(r_{\Sigma}-r)}{2F_{0}M\cosh^{4}[\frac{\omega}{2}(t-t_{0})]}-
\frac{3}{F_{0}^{2}}\bigg\{\dot{c_{1}}^{2}(t)(r_{\Sigma}-r)^{2}+\\\nonumber&\frac{F_{0}\dot{c_{1}}(t)
(r_{\Sigma}-r)}{M}
\frac{\sinh[\frac{\omega}{2}(t-t_{0})]}{\cosh^{3}[\frac{\omega}{2}(t-t_{0})]}\bigg\}-\frac{2\ddot{c_{1}}(t)
(r_{\Sigma}-r)}{F_{0}^{2}}\bigg\{\frac{F_{0}}{6\omega M}\bigg(3\tanh^{2}[\frac{\omega}{2}(t-t_{0})]-1\bigg)
-c_{1}(t)(r_{\Sigma}-r)\bigg\}\bigg]-\\\nonumber&\frac{1}{8\pi}\bigg[\frac{f-Rf_{R}}
{2}-\frac{1}{F_{0}^{2}}\bigg\{\frac{F_{0}}{6\omega M}\bigg(3\tanh^{2}[\frac{\omega}{2}(t-t_{0})]-1\bigg)
-c_{1}(t)(r_{\Sigma}-r)\bigg\}^{2}\ddot{f_{R}}-\frac{1}{F_{0}^{2}}\bigg\{\frac{F_{0}}{6\omega M}\bigg(3\tanh^{2}[\frac{\omega}{2}(t-t_{0})]-1\bigg)-\\\nonumber&c_{1}(t)(r_{\Sigma}-r)\bigg\}
\bigg\{\frac{-F_{0}}{2M}\frac{\cosh[\frac{\omega}{2}(t-t_{0})]}{\sinh^{3}[\frac{\omega}{2}
(t-t_{0})]}-\dot{c_{1}}(t)(r_{\Sigma}-r)\bigg\}\dot{f_{R}}+3c_{1}(t)\bigg\{\frac{F_{0}}{6\omega M}\\\label{76}&\bigg(3\tanh^{2}[\frac{\omega}{2}(t-t_{0})]-1\bigg)-c_{1}(t)(r_{\Sigma}-r)\bigg\}f'_{R}\bigg],\\
\nonumber
P_{\bot}&=\frac{f_{R}}{8\pi}\bigg[3c_{1}^{2}(t)+\frac{[F_{0}-3\omega c_{1}M(r_{\Sigma}-r)]}{3F_{0}M^{2}\cosh^{2}[\frac{\omega}{2}(t-t_{0})]}+\frac{[F_{0}-6\omega c_{1}M(r_{\Sigma}-r)]}{4F_{0}M^{2}\cosh^{4}[\frac{\omega}{2}(t-t_{0})]}-\frac{3}{F_{0}^{2}}\bigg\{\dot{c_{1}^{2}}(t)
(r_{\Sigma}-r)^{2}+\frac{F_{0}\dot{c_{1}}(t)(r_{\Sigma}-r)}{M}\\\nonumber&\frac{\cosh[\frac{\omega}{2}(t-t_{0})]}{\cosh^{3}
[\frac{\omega}{2}(t-t_{0})]}\bigg\}-\frac{3}{F_{0}^{2}}\bigg\{\frac{F_{0}^{2}}{4M^{2}}\frac{\sinh^{2}[\frac{\omega}
{2}(t-t_{0})]}{\cosh^{6}[\frac{\omega}{2}(t-t_{0})]}+\dot{c_{1}^{2}}(t)(r_{\Sigma}-r)^{2}+\frac{F_{0}\dot{c_{1}}(t)
(r_{\Sigma}-r)}{M}\frac{\sinh[\frac{\omega}{2}(t-t_{0})]}{\cosh^{3}[\frac{\omega}{2}(t-t_{0})]}\bigg\}\bigg]
\\\nonumber&-\frac{1}{8\pi}\bigg[\frac{f-Rf_{R}}{2}+\bigg\{\frac{F_{0}}{6\omega M}\bigg(3\tanh^{2}[\frac{\omega}{2}(t-t_{0})]-1\bigg)-c_{1}(t)(r_{\Sigma}-r)\bigg\}\bigg\{\frac{1}{F_{0}^{2}}
\ddot{f_{R}}+f''_{R}\bigg\}\\\nonumber&+\bigg\{\frac{F_{0}}{6\omega M}\bigg(3\tanh^{2}[\frac{\omega}{2}(t-t_{0})]-1\bigg)
-c_{1}(t)(r_{\Sigma}-r)\bigg\}\bigg\{\frac{1}{F_{0}^{2}}\bigg(\frac{-F_{0}}{2M}\frac{\sinh[\frac{\omega}{2}(t-t_{0})]}
{\cosh^{3}[\frac{\omega}{2}(t-t_{0})]}-\dot{c_{1}}(t)(r_{\Sigma}-r)\bigg)\dot{f_{R}}-
\\\label{77}&c_{1}(t)f'_{R}\bigg\}\bigg].
\end{align}
The above-mentioned model described the matter variables of the non-radiating spherical anisotropic objects in an environment of $\chi_{\upsilon}V^{\upsilon}=0$.

\end{document}